\documentclass[reprint]{revtex4-2}

\usepackage{graphicx}
\usepackage{hyperref}
\usepackage{dcolumn}
\usepackage{bm}
\usepackage{braket}
\usepackage{amsmath}
\usepackage{booktabs}
\usepackage{subcaption}
\usepackage{tabularx}
\usepackage{algorithm}
\usepackage{amssymb}
\usepackage{algpseudocode}
\usepackage{subcaption}
\usepackage[dvipsnames]{xcolor}

\usepackage{ulem}

\begin{document}

\preprint{APS/123-QED}

\title{Quantum parameter estimation with uncertainty quantification from continuous measurement data using neural network ensembles}

\author{Amanuel Anteneh}
\email{asa2rc@virginia.edu}
\affiliation{Department of Physics, University of Virginia, 382 McCormick Rd, Charlottesville, VA 22903, USA}

\date{\today}

\begin{abstract}
We show that ensembles of deep neural networks, called deep ensembles, can be used to perform quantum parameter estimation while also providing a means for quantifying uncertainty in parameter estimates, which is a key advantage of using Bayesian inference for parameter estimation that is lost when using existing machine learning methods. 
We show that optimizing for both accurate parameter estimates and well calibrated uncertainty estimates does not lead to degradation in the former as opposed to only optimizing for accuracy. 
We also show that the drift detection capabilities of these ensemble models can be used to detect drift in the experimental data used during inference.
This approach is also shown to provide much faster inference time than both likelihood-based and likelihood-free Bayesian inference.
These results suggest that such models could enable accurate, real-time parameter estimation with quantified uncertainty, making them promising candidates for deployment in experimental settings. 
\end{abstract}

\maketitle

\section{Introduction}
The precise estimation of physical parameters is of central importance in quantum information science and physics more broadly  \cite{kok2010introduction, wiseman2009quantum}. It is particularly relevant to the field of quantum metrology which itself has been instrumental for making discoveries in fundamental physics. 
For example, squeezed-state based metrological methods have been used for gravitational wave detection \cite{yu2020quantum, tse2019quantum, aasi2013enhanced} and dark matter axion searches \cite{backes2021quantum, brady2022entangled, shi2023ultimate}. 
Quantum parameter estimation procedures generally have three steps: i) a quantum probe system is prepared in some initial state $\rho_0$; ii) the unknown parameter(s) $\theta$ are encoded into the state of the probe system using some unitary transformation $U(\theta)\rho_0U^\dag(\theta)$; iii) a measurement is performed on the probe system and the measurement result is used to extract information about the unknown parameter(s). This process is then repeated multiple times to gather more measurement data to improve the precision of the estimation \cite{kok2010introduction}. 

A popular approach for performing quantum parameter estimation is based on Bayesian inference \cite{kiilerich2014estimation, gammelmark2013bayesian, kiilerich2016bayesian}.
This approach provides benefits such as uncertainty quantification for parameter estimates, a natural mechanism by which one can incorporate prior knowledge about the parameter into the analysis, and, perhaps most importantly, the ability to provide an optimal estimator of the parameters under consideration in the limit of a large enough number of measurements \cite{gammelmark2013bayesian, kiilerich2016bayesian}.
However, this approach also comes with significant drawbacks. 

Firstly, Bayesian inference protocols generally require derivation of an analytic expression of the likelihood distribution that the measurement data is assumed to be generated from. 
This is not always feasible for many systems under ideal conditions, such as the high dimensional optomechanical system studied in Refs.~\cite{clark2022exploiting, clark2025efficient} in the non-linear regime of strong optomechanical coupling, and for simple systems this can become difficult when one attempts to model the effects of experimental imperfections such as the case of finite detector efficient for the coherently driven two-level atom studied in~\cite{kiilerich2014estimation}.
This may be circumvented by use of so-called `likelihood-free' methods such as approximate Bayesian computation (ABC) \cite{sisson2018handbook, clark2025efficient} but this method presents it's own challenges such as a drastic increase in sample complexity when the number of parameters being estimated increases, requiring access to a simulator which can generate samples from the likelihood distribution, and selection of a suitable discrepancy measure (and summary statistic) for the rejection sampling routine \cite{murphy2023probabilistic, gelman1995bayesian, sisson2018handbook}. 

Secondly, Bayesian inference time can scale poorly as the dimensionality of the problem, i.e. number of parameters being estimated, increases. 
This is due to the random walk behavior used by common Markov chain Monte Carlo (MCMC) methods, such as Gibbs sampling, to explore the posterior distribution. 
This can lead to long convergence times when the posterior distribution is high dimensional \cite{gelman1995bayesian}. 
While more sophisticated Bayesian inference methods based on the Hamiltonian Monte Carlo algorithm are more robust to this they also introduce the additional overhead of requiring gradient computations to more effectively explore the posterior distribution \cite{lambert2018student, gelman1995bayesian}.
Bayesian inference time can also scale poorly when the number of measurement results used is large \cite{gawlikowski2023survey}. For these reasons real time estimation is made difficult.

\begin{figure*}[ht]
    \centering
    \includegraphics[width=\textwidth]{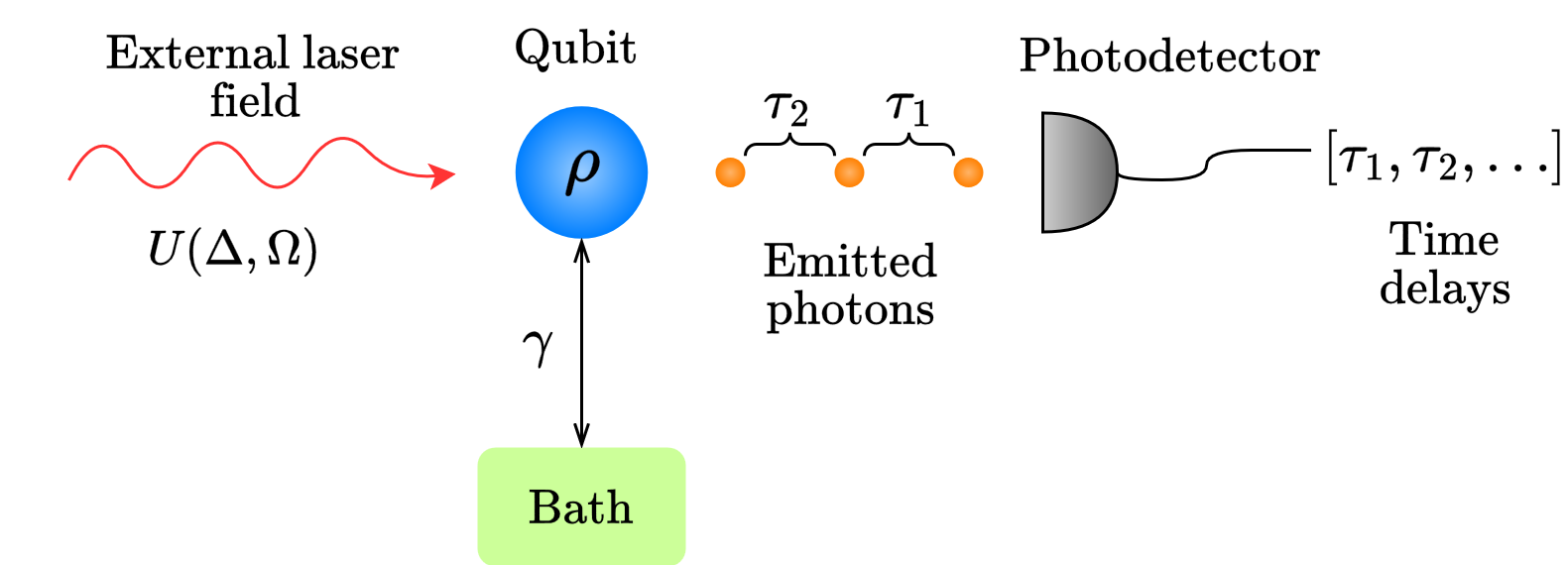}
    \caption{Diagram of quantum parameter estimation procedure using continuous photon counting measurement of a two-level system (qubit) coupled to an external bath (the environment) with coupling strength $\gamma$ and continuously driven by an external laser field. }
    \label{fig:tls_diagram}
\end{figure*}

Recently approaches based on the use of machine learning (ML) algorithms, in particular neural networks (NNs), have gained popularity for performing quantum parameter estimation \cite{greplova2017quantum, khanahmadi2021time, nolan2021machine, rinaldi2024parameter, genois2021quantum, cimini2023deep}. In contrast to Bayesian inference, NN based approaches provide faster inference time, owing to their highly parallelizable inference process, while retaining the benefits of likelihood-free methods like ABC as they can learn complex functions from input-output pairs of training examples due to their high expressive power \cite{cybenko1989approximation}. 
Furthermore, the ABC-based estimation method presented in Ref.~\cite{clark2025efficient} requires the \textit{a priori} preparation of multiple sets of measurement results for different values of $\theta$ to perform inference. 
While an ML approach also requires this for the initial model training phase it is typically no longer needed at inference time unlike the ABC method.
In Ref. \cite{rinaldi2024parameter} it was shown that a NN can be trained to perform parameter estimation given input-output pairs of quantum measurement results from a continuous measurement of a quantum probe system and the corresponding parameter values which generated those measurement results. 
This approach however utilized a large amount of training data and only provides point estimates without uncertainty quantification. 
Here we extend this approach to provide uncertainty estimates, thereby retaining a benefit of Bayesian approaches, and also drastically reducing the amount of training data required to train the model. We also study the robustness of our method to different types of noise in the training data and measurement results provided to the model at inference time.

\section{Quantum System}

As a specific example we study the system from Refs. \cite{gammelmark2013bayesian, kiilerich2014estimation, gammelmark2014fisher, rinaldi2024parameter, clark2025efficient}, shown in Fig. \ref{fig:tls_diagram}, which consists of a two-level system (TLS) with states $\{\ket{g},\ket{e}\}$ and transition frequency $\omega_q$ which is continuously driven by an external laser field with frequency $\omega_L$. The TLS is also coupled to the environment which results in dissipative interactions that give rise to stochastic photon emission with rate $\gamma$.   
The dynamics of this system can be described by the Lindblad master equation  \cite{kiilerich2014estimation} (where $\hbar=1$)

\begin{align}{\label{master_eq}}
\frac{\partial }{\partial t}\hat{\rho}(t) &= -i[\hat{H},\hat{\rho}(t)] +  
\gamma\hat{L}\hat{\rho}(t)\hat{L}^\dag - \frac{\gamma}{2}\{\hat{L}^\dag\hat{L}, \hat{\rho}(t)\}
\end{align}
where, in the reference frame rotating at the frequency of the laser, the Hamiltonian $\hat{H}$ can be written as

\begin{align}
    \hat{H} = \Delta\hat{\sigma}_{_{\scriptscriptstyle -}}^\dag\hat{\sigma}_{_{\scriptscriptstyle -}} + \Omega(\hat{\sigma}_{_{\scriptscriptstyle -}} + \hat{\sigma}_{_{\scriptscriptstyle -}}^\dag).
\end{align}
Here $\Delta \equiv \omega_q-\omega_L$ and $\Omega$ are the laser-atom detuning and the Rabi frequency respectively, $\gamma$ is the rate at which the excited state of the TLS decays to the ground state $\ket{g}$, thus emitting a photon, $\hat{\sigma}_{_{\scriptscriptstyle -}} = \ket{g}\bra{e}$ is the TLS lowering operator and the Lindblad jump operator is set as $\hat{L} = \hat{\sigma}_{_{\scriptscriptstyle -}}$. 

The first term in Eq. \ref{master_eq} corresponds to the unitary dynamics of the system governed by the Hamiltonian $\hat{H}$ while the last two terms model the dissipative dynamics of the TLS's interaction with the environment. 
The second term corresponds to the collapse of the TLS to the ground state which coincides with the measurement of a spontaneously emitted photon. The anti-commutator term corresponds to the dissipative dynamics in the absence of photon detection \cite{kiilerich2014estimation}. It was first shown in Ref. \cite{kiilerich2014estimation} that measurements of the time delays between consecutive photon detections on the system could be used to estimate the parameters $\Delta$ and $\Omega$ using Bayesian inference.

\section{Machine Learning Experiments}{\label{ml_exp}}

We train and test our models for predicting the detuning parameter $\Delta$ with the same dataset used in \cite{rinaldi2024parameter}. As such the inputs to our NNs are a set of time delays $x=[\tau_1, ...,\tau_N]$ where $N=48$ which were generated using the Monte Carlo method of quantum trajectories with the \texttt{QuTiP} library \cite{johansson2012qutip}. The decay rate is assumed to be known at $\gamma=1$ and that $\Omega=\gamma$.
All models were implemented using the \texttt{PyTorch} library and we utilized the automatic mixed precision training method provided by \texttt{PyTorch} to reduce memory overhead \cite{narang2017mixed, paszke2019pytorch}.
All training ran on an NVIDIA GeForce RTX 2070 Super GPU. 

\begin{figure*}[ht]
    \centering
    \includegraphics[scale=0.2]{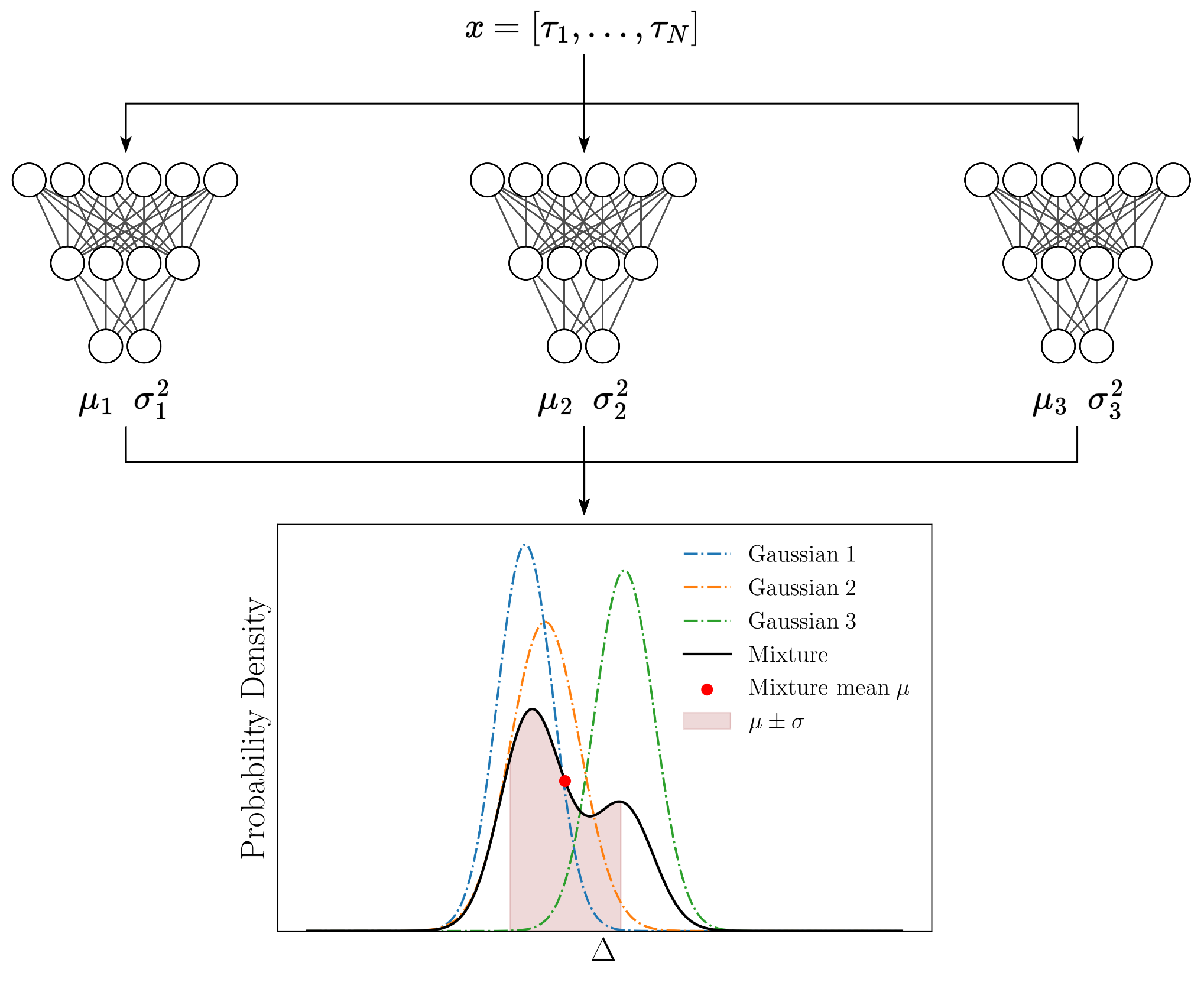}
    \caption{Diagram of using deep ensemble of $M=3$ NNs for estimating the detuning parameter $\Delta$ from a set of input time delay measurements $x=[\tau_1,\dots, \tau_N]$ from a quantum trajectory. }
    \label{fig:ensemble_diagram}
\end{figure*}

\subsection{Deep ensembles}
Model ensembling is a simple but powerful method in machine learning. Ensemble methods employ multiple models, typically ‘weak learners’ as individually these models have suboptimal performance, to create a single overall model by aggregating the predictions of the component models. The resulting model often has better performance, in terms of both bias and variance, than the individual models in isolation \cite{hastie2009elements}. 
A state-of-the-art method for predictive uncertainty quantification for deep learning based models are deep ensembles \cite{lakshminarayanan2017simple, ovadia2019can, gawlikowski2023survey}. 
These models quantify epistemic (model)  uncertainty, i.e. uncertainty that arises from the choice of model parameters, due to the use of of an ensemble of independent models.
This type of uncertainty is reducible if one can procure more data.
They also quantify aleatoric (data) uncertainty, i.e. uncertainty that is inherent to the process which produces the training data, by being trained to minimize a negative log likelihood loss function~\cite{gawlikowski2023survey, thuy2025learning, thuy2024explainability}.
Unlike model uncertainty, data uncertainty is not reducible~\cite{thuy2025learning}.
For regression problems, such as the one considered in this work, deep ensembles work by training $M$ independent NNs to learn a continuous probability distribution over predictions. The overall model is then treated as a mixture of these distributions.
Model diversity emerges from the random initialization of each network at the start of training as well as the randomness from the mini-batch sampling that occurs during training via stochastic gradient descent.  

Deep ensembles are preferable to Bayesian NNs (BNNs) \cite{murphy2023probabilistic}, which are another common tool for quantifying uncertainty in deep learning, for two reasons. Firstly they require minimal changes to the standard NN training process and their training is easily scalable via parallelization since each network in the ensemble can be trained completely independently. Secondly BNNs face the same computational problem as Bayesian inference in general i.e. the intractability of computing the posterior distribution over model parameters exactly \cite{murphy2023probabilistic, lakshminarayanan2017simple}. 
In fact due to the large number of parameters in NNs standard Bayesian sampling methods like MCMC are often too computationally expensive to use for BNN training and methods such as variational inference, which are faster but have weaker theoretical guarantees compared to MCMC, must be used \cite{gawlikowski2023survey}.  

\subsection{Model training}
We benchmark our deep ensembles against Bayesian inference as well as single NNs. 
All NNs use the same architecture, with the exception of the input layer, as the \textit{Hist-Dense} network in Ref. \cite{rinaldi2024parameter} with three hidden layers of size 100, 50, and 30 respectively using ReLU activation functions.
For the input layer of all our models we use a custom layer that implements a smoothed histogram based on ideas from kernel density estimation \cite{hastie2009elements}. 
This introduces an inductive bias into the model i.e. that the ordering of the input time delays has no significance as the TLS is reset to the ground state in the event of a photon detection so the time delay measurements are independent and identically distributed \cite{bishop2023deep}. 
This also allows the NN to process inputs of arbitrary length which is useful for when varying numbers of time delay measurements are made \cite{rinaldi2024parameter}.
We use equally spaced bins where the density at bin $b$ is given by a sum over all values in the sample in the form of the Gaussian kernel
\begin{align}
    f(b) = \sum_{i=0}^N\exp{\left( -\frac{(\tau_i - b)^2}{2\phi^2} \right)}
\end{align}
where $\phi^2$, sometimes called the bandwidth, is a learnable parameter which controls the smoothness of the resulting estimated kernel density and we use the same range for the histogram, $\tau_{\min}=0$ and $\tau_{\max}=100/\gamma$, as in \cite{rinaldi2024parameter}.  We treat the number of bins as a tunable hyperparameter.

Using the deep ensemble we seek to learn a distribution  $P(\Delta|\tau_1,...,\tau_N)$ over values of $\Delta$ given the time delay measurements from a quantum trajectory as shown schematically in Fig. \ref{fig:ensemble_diagram}. We approximate this distribution as an evenly weighted mixture of $M=10$ Gaussian distributions and have the $m$-th NN in the ensemble learn the mean, $\mu_m$, and variance, $\sigma^2_m$, of the $m$-th component distribution by minimizing the Gaussian negative log likelihood loss \cite{lakshminarayanan2017simple}

\begin{align}
\mathcal{L}(y, \mu, \sigma^2) = \frac{1}{2} \log(\sigma^2) + \frac{(y - \mu)^2}{2\sigma^2} + \textrm{constant}
\end{align}
where $y=\Delta$ is the ground truth detuning parameter value for the trajectory-parameter pair $([\tau_1,...,\tau_N], \Delta)$.
The prediction for the parameter is then taken to be the mean of the mixture while the predictive uncertainty is quantified by the variance of the mixture. These quantities are given by
\begin{align}
    \mu(x) &= \frac{1}{M} \sum_{m} \mu_m(x) \\
    \sigma^2(x) &= \frac{1}{M} \sum_{m}\left( \sigma_m^2(x) + \mu_m^2(x)\right) - \mu^2(x)
\end{align}
respectively \cite{lakshminarayanan2017simple}.
The total uncertainty $\sigma^2(x)$ can be written in terms of aleatoric and epistemic uncertainty as
\begin{align}
\sigma^2(x)=\underbrace{\frac1M\sum_m\sigma_m^2(x)}_{\text{aleatoric } \sigma_{\large \textrm{al}}^2}+\underbrace{\frac1M\sum_m\mu_m^2(x)-\left(\frac1M\sum_m\mu_m(x)\right)^2}_ {\text{epistemic } \sigma_{\large \textrm{ep}}^2}.
\end{align}
A useful property of Gaussian mixtures is that they are universal approximators for probability densities~\cite{goodfellow2016deep}.
Note, however, that the log likelihood need not be that of a Gaussian, the negative log likelihood of any probability distribution may be used if deemed appropriate and one is able to compute the relevant moments. 

An alternative approach to training an ensemble of $M$ NNs would be to simply train each network to minimize the mean squared error (MSE) loss function. 
The uncertainty estimates would then be computed as the empirical variance of the predictions made by the $M$ networks. While this approach is more straightforward it leads to the model producing highly overconfident uncertainty estimates as MSE only optimizes for predictive accuracy and not predictive uncertainty \cite{lakshminarayanan2017simple}. This highlights the crucial role played by selection of a proper loss function like the Gaussian negative log likelihood in training a deep ensemble to produce properly calibrated uncertainty estimates. 

\subsection{Hyperparameter Tuning}
Ref. \cite{rinaldi2024parameter} trained single NNs on a dataset of $4 \times 10^6$ trajectory-parameter pairs \cite{sanchez_munoz_2023_8305509}. With the use of hyperparameter tuning we are able to achieve the same level of performance with only $1\%$ ($4 \times 10^4$) of the original training data. 
This is also much lower than the amount of data needed for ABC which used 1000 trajectories generated for 100 different values of $\Delta$ which results in a total dataset size of $10^5$ \cite{clark2025efficient, 4AC59W_2025}. 
Note also that the ABC approach used longer trajectories with $N=200$ time delay measurements as opposed to the $N=48$ used in this work.
We split the training data into training and validation sets with a $80\%$ and  $20\%$ split respectively. 
A commonly used approach to train ensembles models is the use of bootstrap aggregation (bagging) which consists of training ensemble members on different bootstrapped samples of the original dataset. 
However, this is mainly advantageous for ensemble models comprised of weak learners, such as tree-based ensemble models, and worsens performance when used to training ensembles of strong learners like deep neural networks \cite{lee2015m, lakshminarayanan2017simple}. 
We therefore did not utilize bagging in our experiments. 

We use the tree-structured Parzen estimator optimization algorithm provided by the \texttt{Optuna} framework to perform hyperparameter tuning of the single network models presented in this work \cite{akiba2019optuna}. 
We utilize both trail pruning and early stopping during our tuning and training processes respectively and tune for 20 trials. 
For the single models the loss function used during training was also tuned as part of the hyperparameter tuning process with the possible choices being the root mean squared error (RMSE) loss or the mean squared logarithmic error (MSLE) loss \cite{rinaldi2024parameter}. 
Further details on the tuned hyperparemeters can be found in Appendix \ref{appendix:B}.
For experiment tracking, validation metric logging, and model storage we utilize the open source framework \texttt{MLflow} \cite{zaharia2018accelerating}. 
We make the \texttt{MLflow} logs which contain learning curve plots, model weights, and hyperparameter settings for all the models trained in the present work available at Zenodo
\cite{anteneh_2025_17014659}. 

\subsection{Predictive accuracy preservation under ensembling}

In this section we examine the performance of our ensemble approach based on predictive accuracy. This is crucial as we must ensure that the training of the ensemble does not sacrifice predictive accuracy in favor of well calibrated uncertainty estimates.

\subsubsection{Noiseless training data}
\begin{figure}
    \centering
    \begin{subfigure}[b]{0.475\textwidth}
    \centering
    \includegraphics[width=1.025\columnwidth]{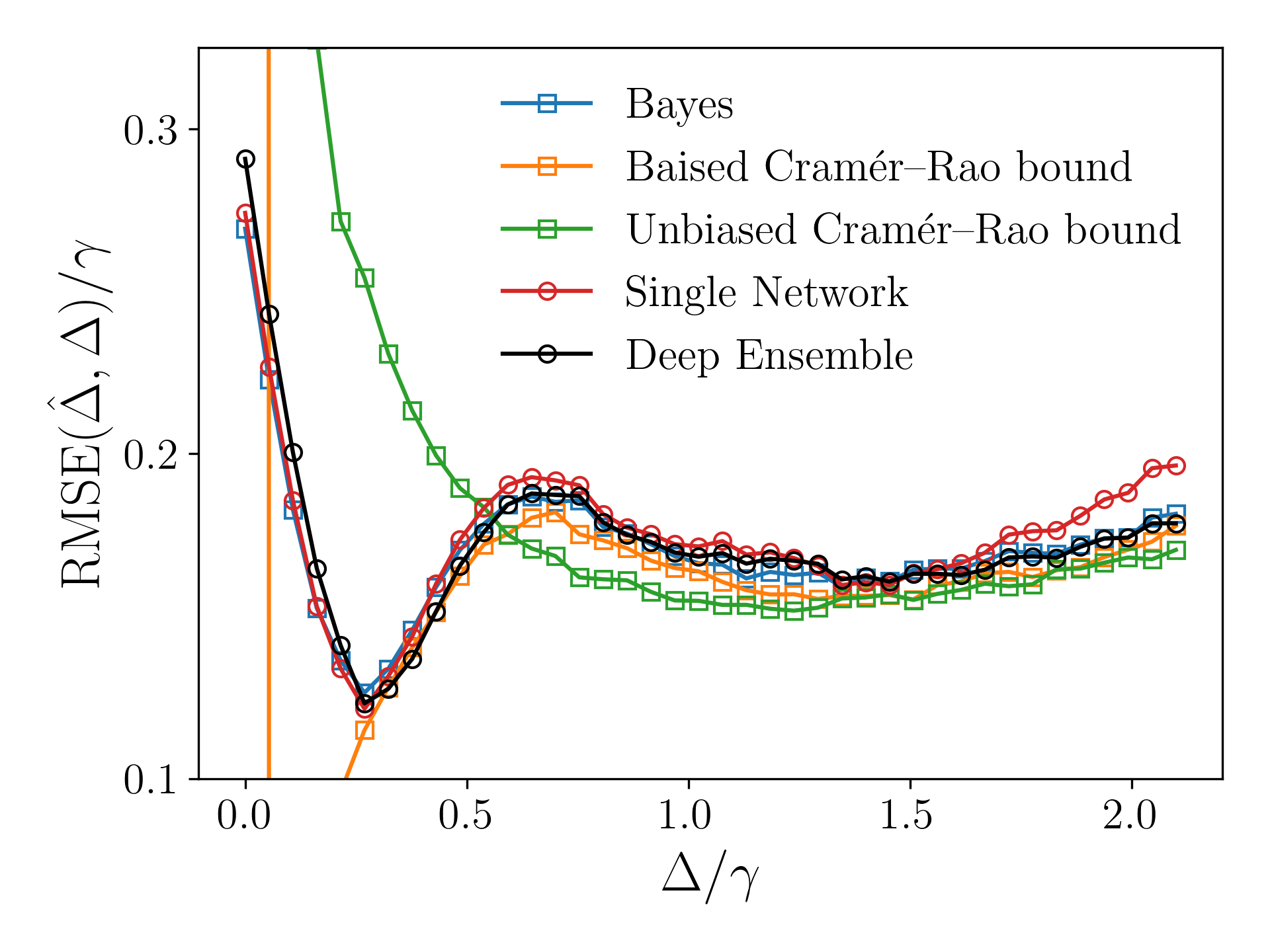}
    \caption{Average RMSE of various estimation strategies plotted with the biased CRB.}
    \label{fig:rmse1D}
    \end{subfigure}
    \begin{subfigure}[b]{0.475\textwidth}
    \centering
    \includegraphics[width=1.025\columnwidth]{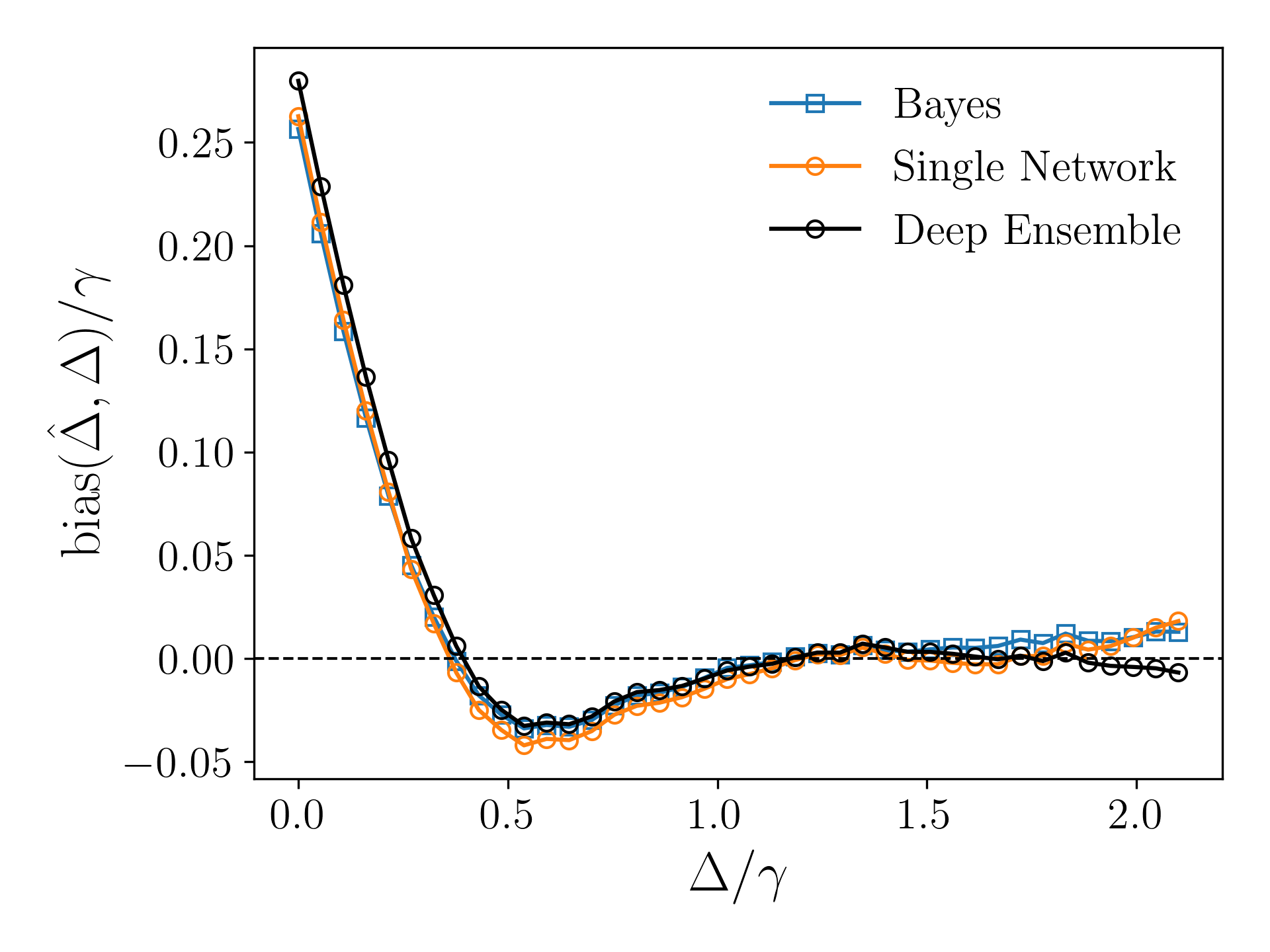}
    \caption{Bias of various estimation strategies. }
     \label{fig:bias1D}
    \end{subfigure}
    \caption{Performance of different estimators on $\Delta$ estimation task.}
\end{figure}

We first benchmark the deep ensemble against Bayesian inference and single NNs in the case of noiseless training data. 
Fig. \ref{fig:rmse1D} shows the biased and unbiased Cramér–Rao bound (CRB) as well as the error of the three estimators on the same test set of $40 \times 10^4$ trajectories generated for 40 values of $\Delta$, which where uniformly chosen over the interval $[0, 2.1\gamma]$, used in \cite{rinaldi2024parameter}. Details on the Bayesian estimation procedure can be found in Appendix \ref{appendix:A} as well as Appendix B of \cite{rinaldi2024parameter}. 
Since all estimators used in this work are biased in addition to the traditional CRB we also plot the biased CRB computed in Ref. \cite{rinaldi2024parameter} which is given by~\cite{barbieri2022optical}
\begin{align}
    \textrm{var}(\hat{\theta}) \geq \left(1 + \frac{d}{d\theta} \textrm{bias}(\hat{\theta}, \theta) \right)^2 \frac{1}{NF(\theta)}
\end{align}
where $N$ is the number of measured time delays in the sample and $F(\theta)$ is the Fisher information~\footnote{Note that the CRBs diverge for $\Delta=0$ as the Fisher information vanishes at this value.}.
The bias of each estimator, defined as $\textrm{bias}(\hat{\Delta}, \Delta)\equiv\mathbb{E}(\hat{\Delta}) - \Delta$ \cite{murphy2022probabilistic}, can be seen in Fig. \ref{fig:bias1D}. 
We can see that the deep ensemble model outperforms or is competitive with the single network model and that the deep ensemble saturates the biased CRB of the Bayesian estimator for more values of $\Delta$ than compared to the single model.
This is particularly notable because, unlike the single model's loss function, the ensemble model’s loss function optimizes not only for predictive error but also for predictive uncertainty. 

\subsubsection{Noise in time delay measurements}
\begin{figure}
    \centering
    \includegraphics[width=1.025\columnwidth]{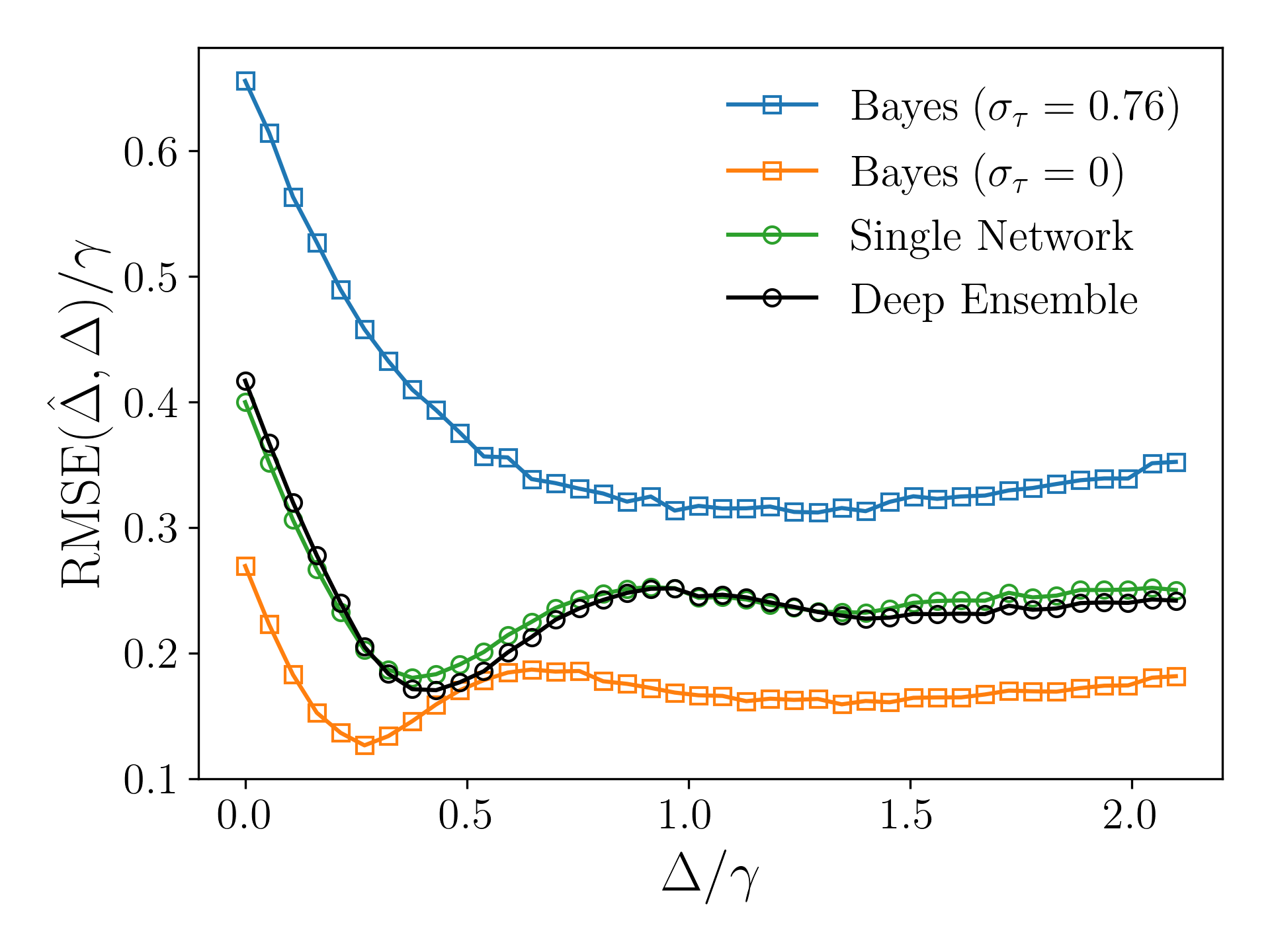}
    \caption{Performance of single network and deep ensemble in the presence of time jitter noise in the training and testing data following a Gaussian distribution $\mathcal{N}(0, \sigma_\tau=0.76)$.}
    \label{fig:x_noise}
\end{figure}

{\label{noise_x_sec}}
For real data from an experimental setup there is likely to be noise from various sources. One such source of noise would be time jitter \cite{rinaldi2024parameter, lopez2022loss}. It was shown in Ref \cite{rinaldi2024parameter} that, as expected, Bayesian inference that does not account for this noise in it's likelihood model performs very poorly and is outperformed by a NN that is trained on data with this noise. 
We evaluate the robustness of our estimator to this noise by modifying the input time delays used for training and testing of our models as $x \rightarrow x + x_{\textrm{noise}}$ where  $x_{\textrm{noise}}$ is a Gaussian noise term $x_{\textrm{noise}} \sim \mathcal{N}(0, \sigma_\tau)$. Fig. \ref{fig:x_noise} shows the results for training data with time jitter noise following the Gaussian distribution $\mathcal{N}(0, \sigma_\tau=0.76)$ with the Bayesian inference results for noiseless and noisy input data shown for reference. 
As expected the Bayesian estimator that does not take into account the time jitter noise performs significantly worse than the noiseless Bayesian estimator. 
We can see that the deep ensemble is again competitive with or outperforms the single model for all values of $\Delta$.

\subsubsection{Noise in ground truth parameter values}

\begin{figure}
    \centering
    \begin{subfigure}[b]{0.475\textwidth}
    \centering
    \includegraphics[width=1.025\columnwidth]{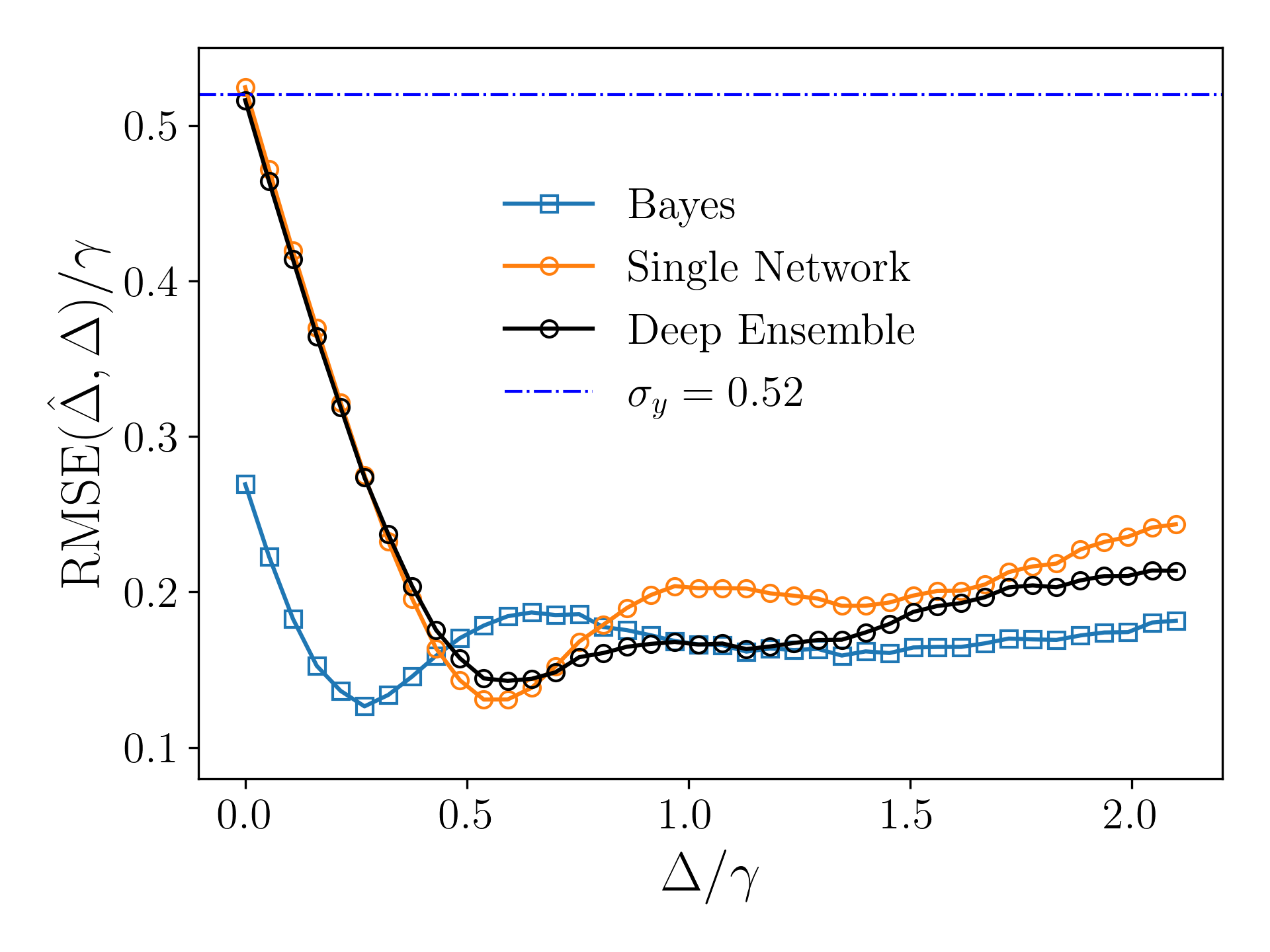}
    \caption{Performance of single network and deep ensemble in the presence of noise in the training data labels following a Gaussian distribution $\mathcal{N}(0, \sigma_y=0.52)$.}
    \label{fig:y_noise}
    \end{subfigure}
    \begin{subfigure}[b]{0.475\textwidth}
    \centering
    \includegraphics[width=1.025\columnwidth]{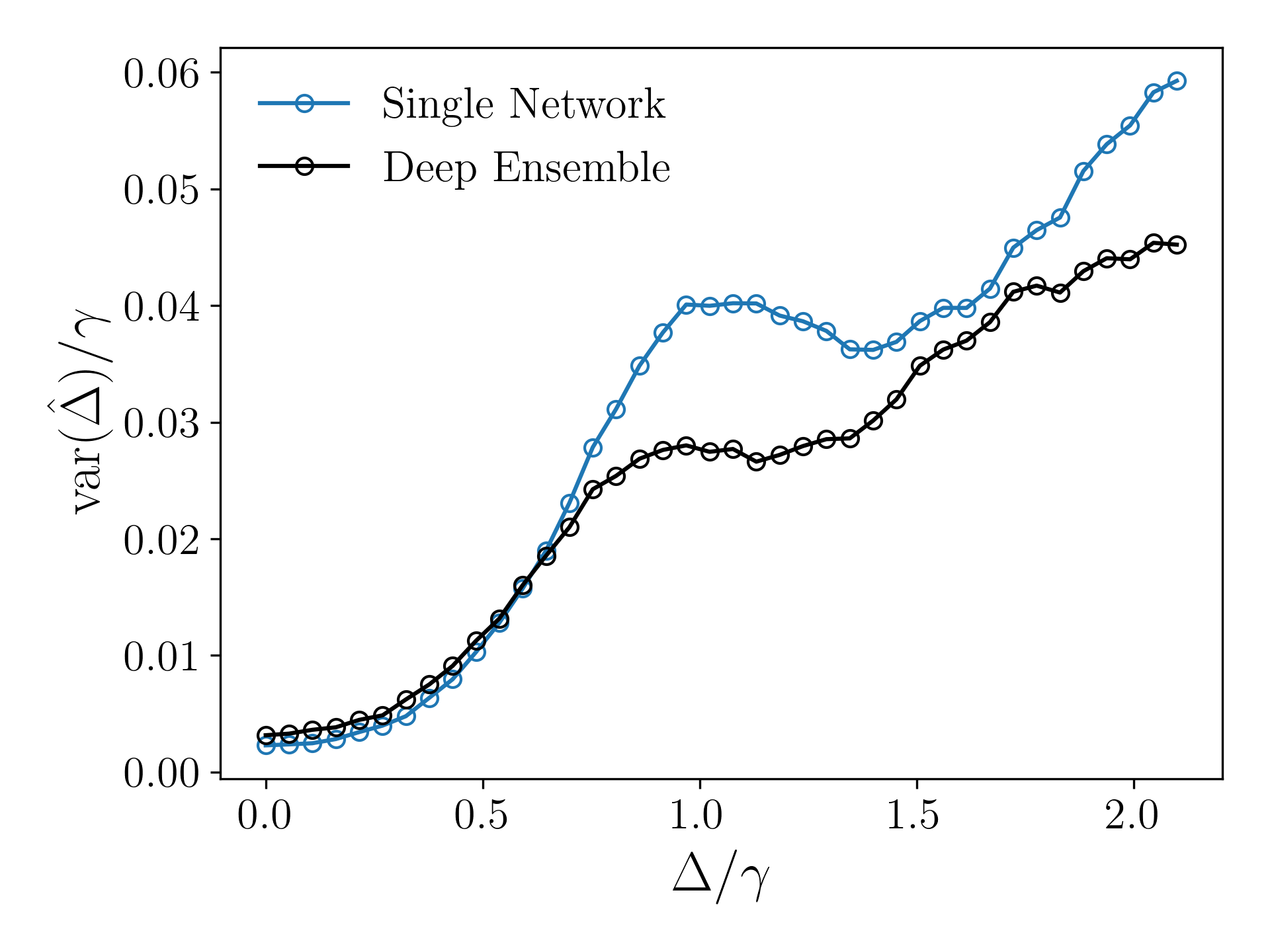}
    \caption{Variance of single network model and deep ensemble trained on data with noisy labels.}
     \label{fig:var_y_noise}
    \end{subfigure}
    \caption{Performance of different estimators on $\Delta$ estimation task with noise present in training labels.}
\end{figure}

Another question addressed in Ref. \cite{rinaldi2024parameter} was whether an NN can learn to perform parameter estimation robustly in the presence of noise in the training labels. 
This could, for example, occur if the method used to generate the training labels, i.e. the measurement device, had less than ideal precision. 
This noise was modeled by adding a Gaussian noise term $\mathcal{N}(0, \sigma_y)$ to the training labels. 
Robustness is defined as being able to make predictions with an RMSE lower than that of the standard deviation $\sigma_y$ of the Gaussian noise in the training labels of $\Delta$. 
It was found that a single NN is indeed robust to such noise \cite{rinaldi2024parameter}.
A natural question would then be whether this robustness is retained when using a deep ensemble.
Fig. \ref{fig:y_noise} answers this question in the affirmative. In fact the deep ensembles achieves better performance than the single model for most values of $\Delta$. 
This increased robustness to perturbations in the training labels is expected as it is a well known strength of ensemble models that they generally have lower variance than the individual ensemble members \cite{james2023statistical}. This is explicitly shown in Fig. \ref{fig:var_y_noise}, where the variance of an estimator is given by $\textrm{var}(\hat{\theta}) \equiv \mathbb{E}[\hat{\theta}^2] - \mathbb{E}^2[\hat{\theta}]$ \cite{murphy2022probabilistic}, which shows that the deep ensemble generally has lower variance than the single network.

\subsection{Uncertainty quantification under data shift}
For ML models which are discriminative models such as the ones used here we assume the input-output pairs, i.e. the data, $(X,Y)$ are generated from some joint distribution $P(X,Y)$ which can be factored into $P(X)P(Y|X)$ \cite{murphy2023probabilistic}. 
An important consideration is how models which are trained on training data generated from one distribution $P(X,Y)$ perform on data generated from a different distribution $Q(X,Y)$ such that $P(X,Y) \neq Q(X,Y)$. 
This phenomena is often seen in ML models deployed in real world settings and is called data shift or data drift, with the data coming from the new distribution being called out-of-distribution samples, and often results in the models performance degrading \cite{quinonero2022dataset}.

\begin{figure}
    \centering
    \begin{subfigure}[b]{0.45\textwidth}
    \centering
    \includegraphics[width=1.025\columnwidth]{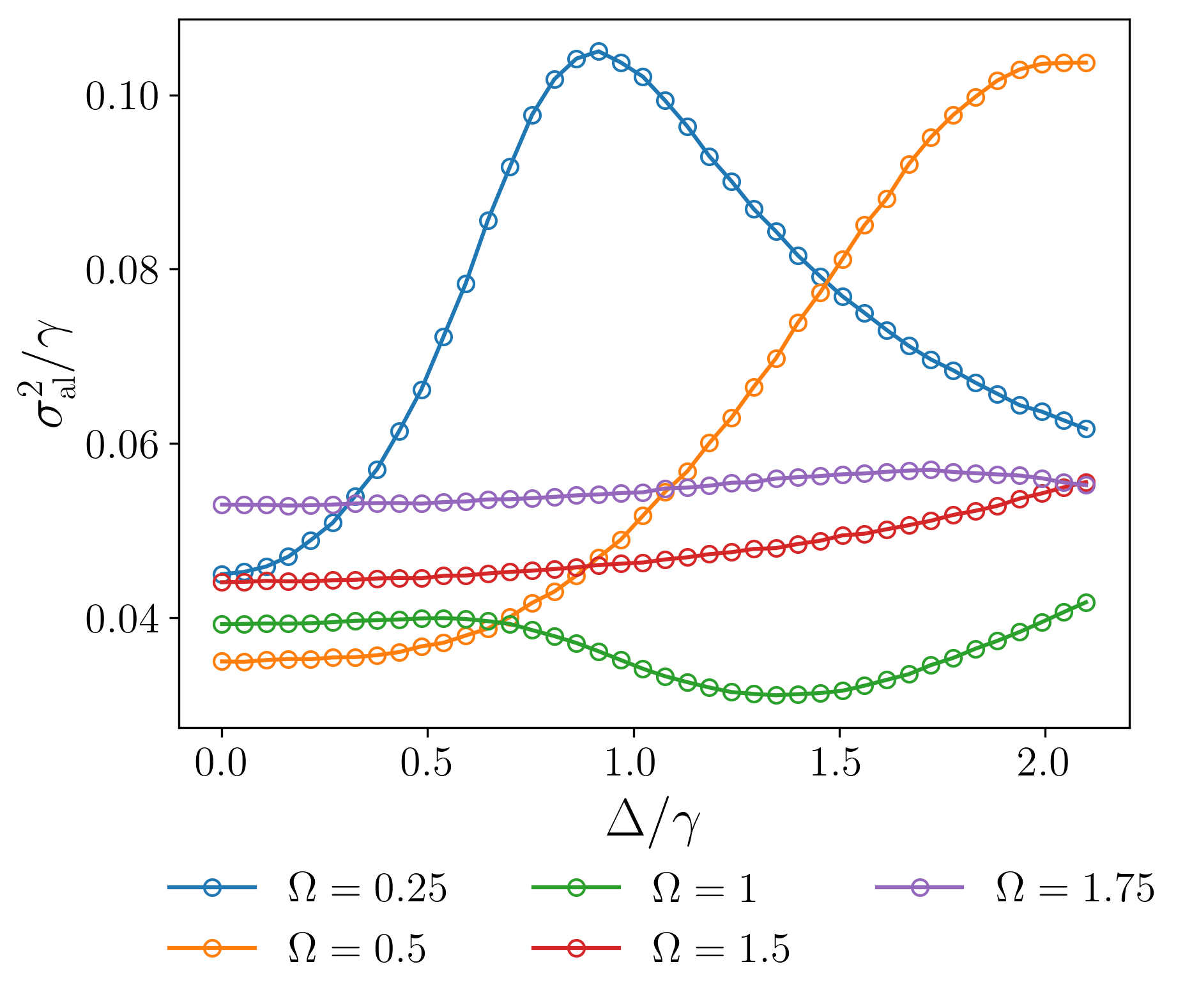}
    \caption{}
    \label{fig:ood_rabi_al}
    \end{subfigure}
    \begin{subfigure}[b]{0.45\textwidth}
    \centering
    \includegraphics[width=1.025\columnwidth]{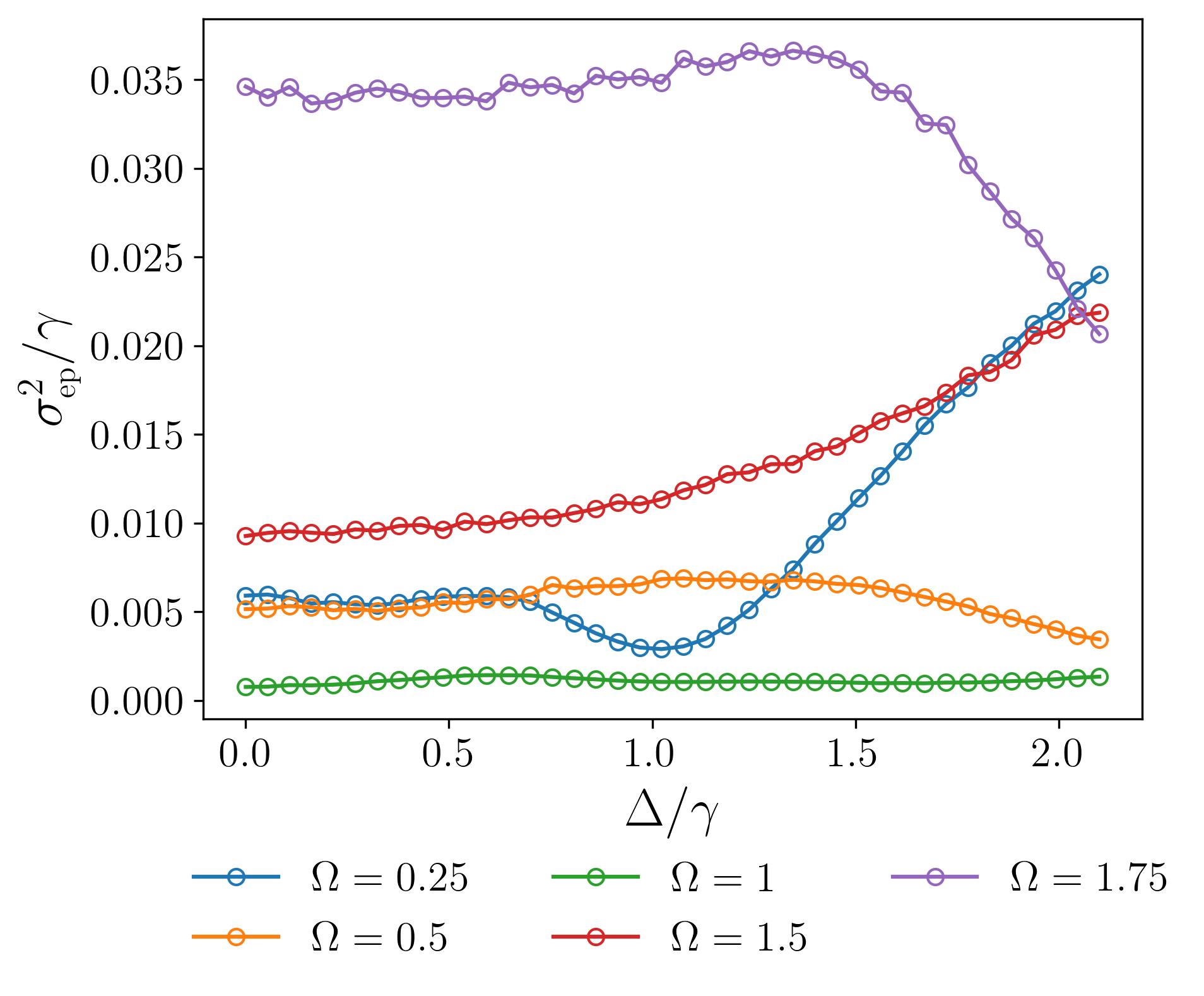}
    \caption{}
    \label{fig:ood_rabi_ep}
    \end{subfigure}
    \begin{subfigure}[b]{0.45\textwidth}
    \centering
    \includegraphics[width=1.025\columnwidth]{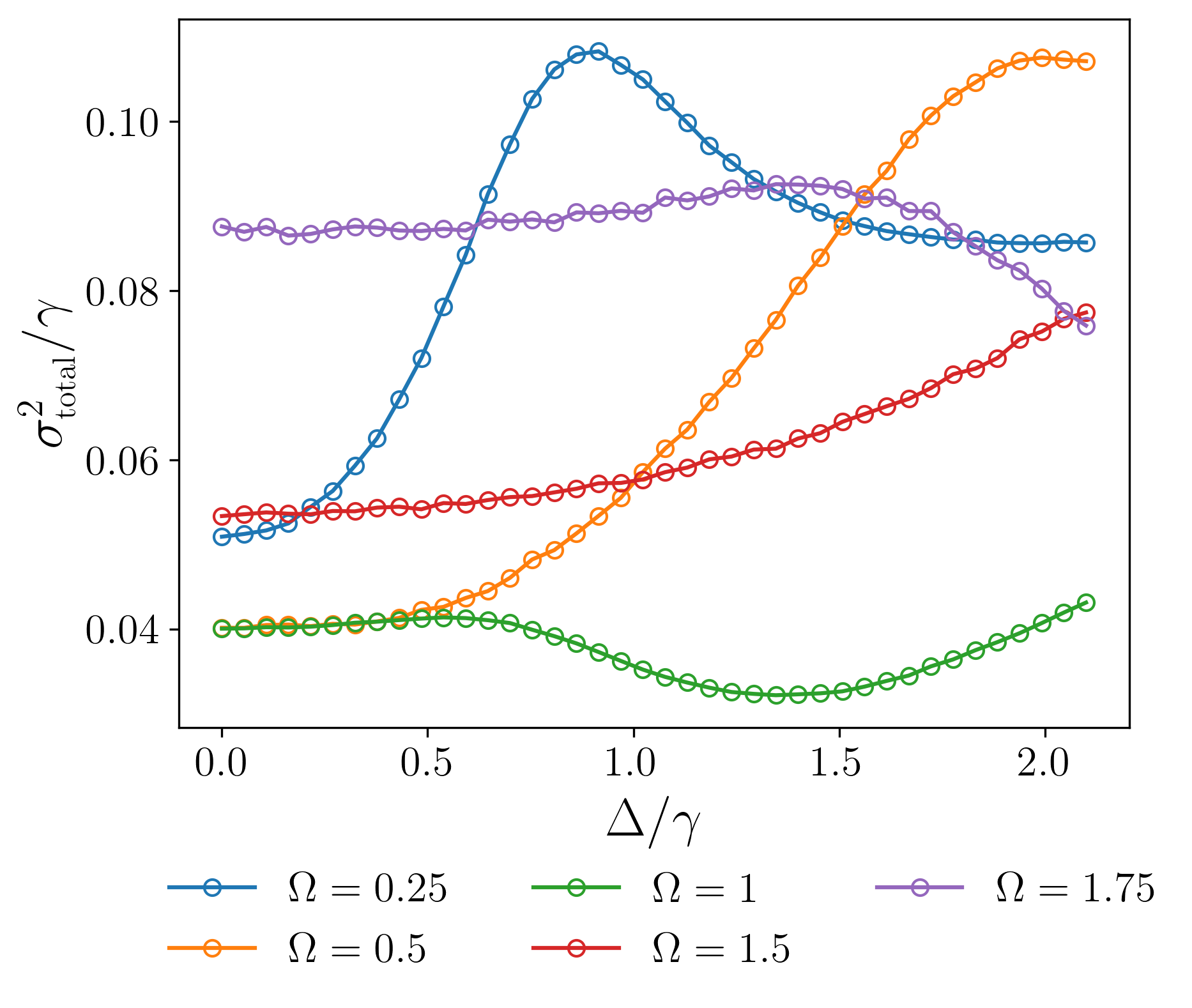}
    \caption{}
    \label{fig:ood_rabi_tot}
    \end{subfigure}
    \caption{Average uncertainty of deep ensemble for test data generated from the training distribution where $\Omega=1$ and distributions where $\Omega \neq 1$. (a) Average aleatoric uncertainty $\sigma_{\large \textrm{al}}^2$. (b) Average epistemic uncertainty $\sigma_{\large \textrm{ep}}^2$. (c) Average total uncertainty $\sigma_{\large \textrm{total}}^2$.}
    \label{fig:ood_rabi}
\end{figure}

An important property to examine for models which provide uncertainty estimates is how the predicted uncertainty changes when the model is provided out-of-distribution examples e.g. in the case of data shift \cite{lakshminarayanan2017simple}. 
Ideally we would want our model to have higher predictive uncertainty when given input data from a distribution that is far from the training distribution i.e. we would expect the predicted Gaussian mixture to have greater variance $\sigma^2$. 
Fig. \ref{fig:ood_rabi} shows the average aleatoric, epistemic and total uncertainty of the deep ensemble model when given sets of time delays from systems in which the Rabi frequency differs from the value used to generate the training data. 
We can see that as the Rabi frequency deviates further from $\Omega=1$ the predictive uncertainty of our model almost always increases for all types of uncertainty with the epistemic uncertainty being the most likely to increase with the data shift.

\begin{figure}
    \centering
    \begin{subfigure}[b]{0.45\textwidth}
    \centering
    \includegraphics[width=1.025\columnwidth]{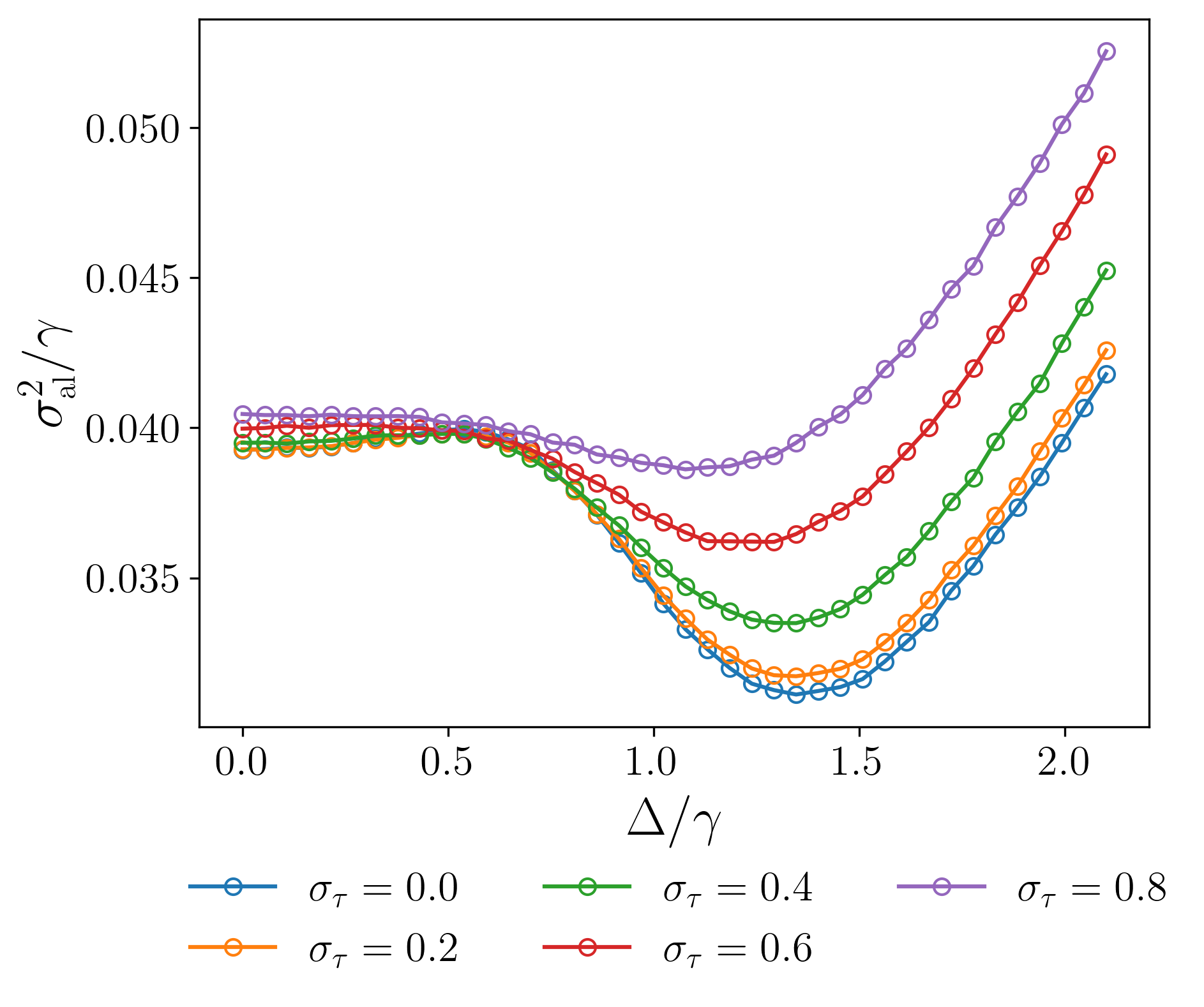}
    \caption{}
    \label{fig:ood_jitter_al}
    \end{subfigure}
    \begin{subfigure}[b]{0.45\textwidth}
    \centering
    \includegraphics[width=1.025\columnwidth]{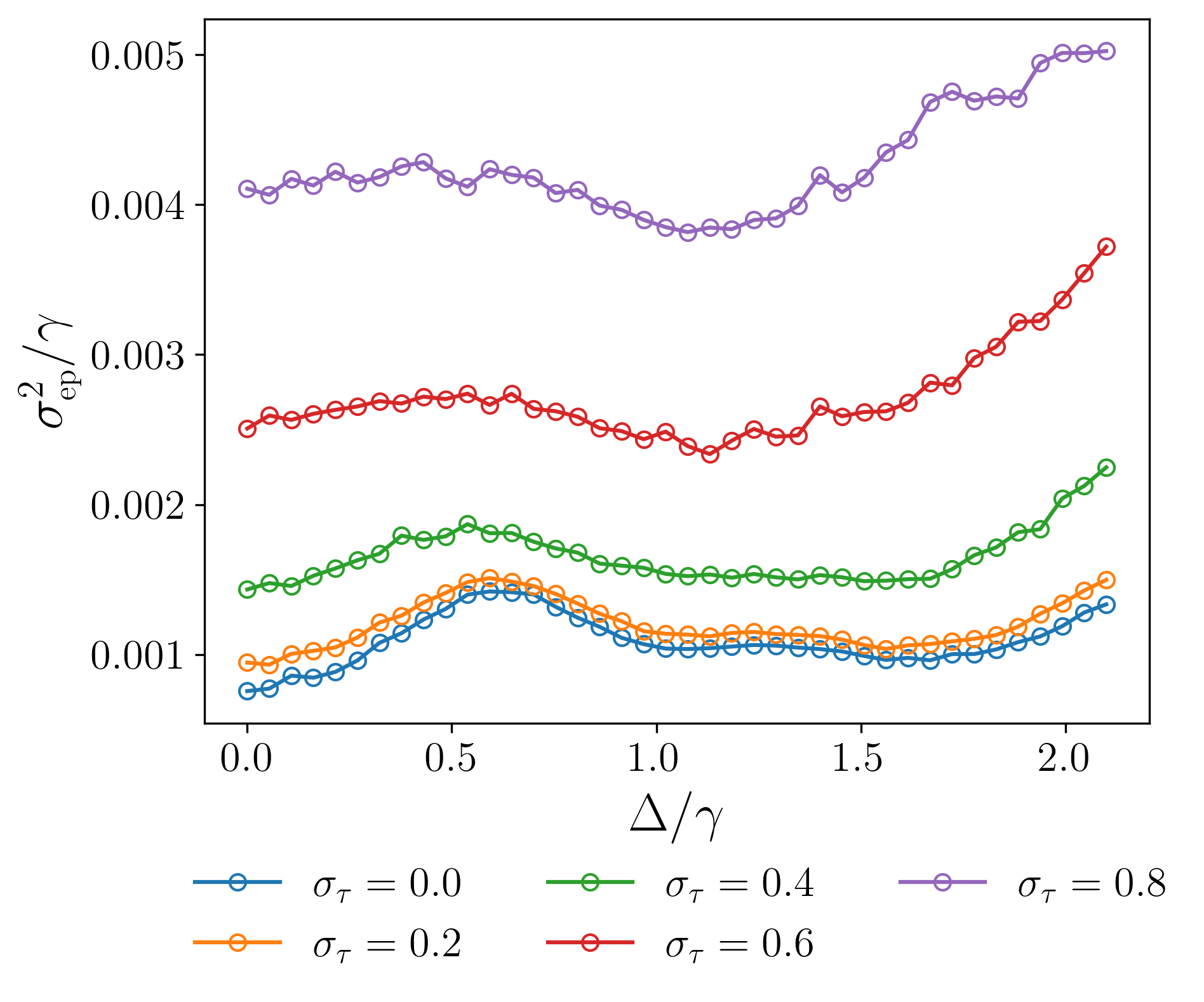}
    \caption{}
    \label{fig:ood_jitter_ep}
    \end{subfigure}
    \begin{subfigure}[b]{0.45\textwidth}
    \centering
    \includegraphics[width=1.025\columnwidth]{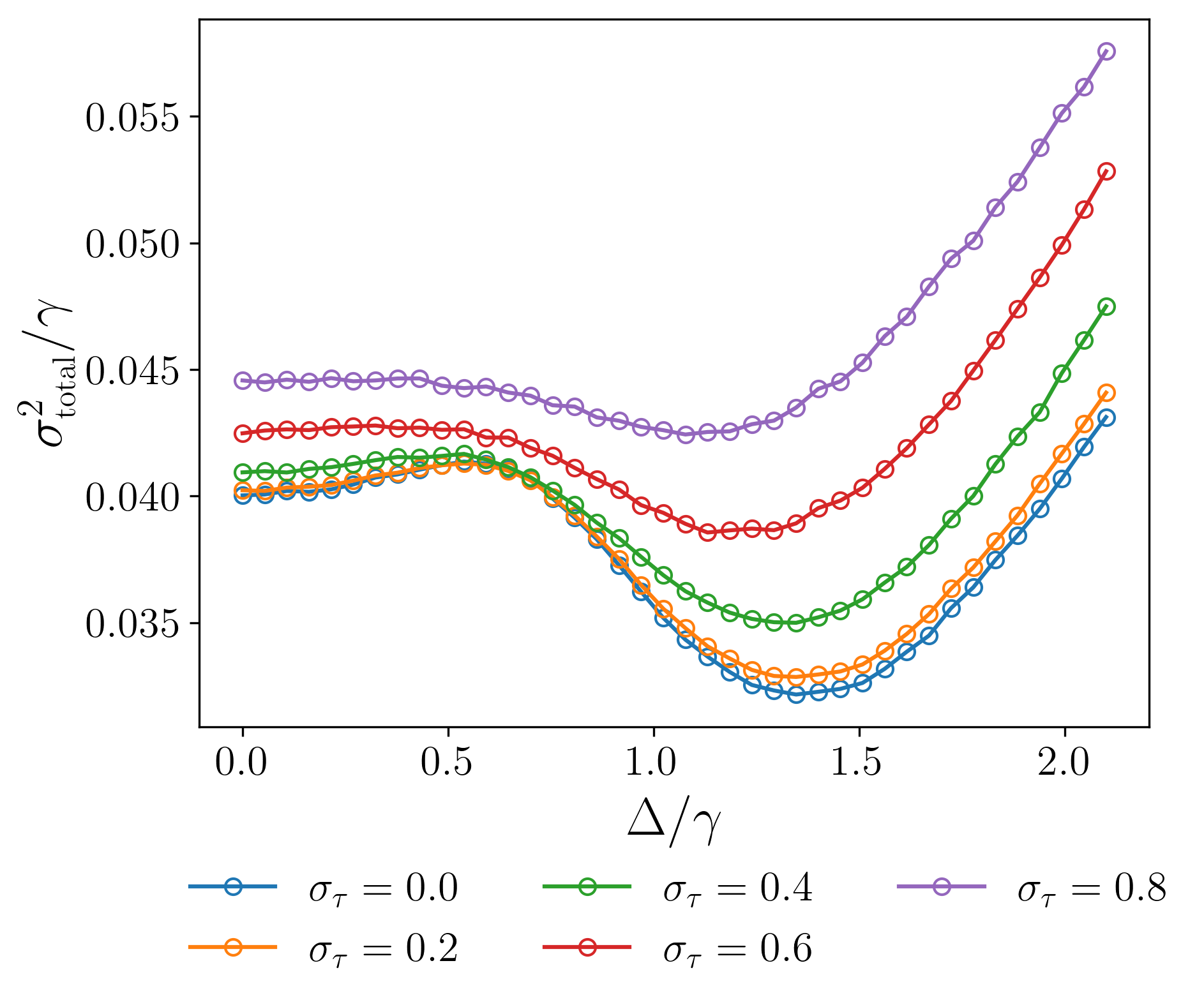}
    \caption{}
    \label{fig:ood_jitter_tot}
    \end{subfigure}
    \caption{Average uncertainty of deep ensemble for test data generated from the training distribution with time jitter noise following the distribution $\mathcal{N}(0, \sigma_{\tau})$. (a) Average aleatoric uncertainty $\sigma_{\large \textrm{al}}^2$. (b) Average epistemic uncertainty $\sigma_{\large \textrm{ep}}^2$. (c) Average total uncertainty $\sigma_{\large \textrm{total}}^2$.}
    \label{fig:ood_jitter}
\end{figure}

This can be used to detect drift in the experimental data at inference time analogous to what has been demonstrated in Ref. \cite{thuy2024fast} for image classification in industrial processes. 
For instance if one trains a deep ensemble on time delay data which were generated using a detector with minimal time jitter noise we would expect the ensemble, at inference time, to produce larger uncertainty estimates when presented with time delay measurements generated using a detector with time jitter noise. 
We can see that this is indeed the case in Fig. \ref{fig:ood_jitter} which shows the average aleatoric, epistemic and total uncertainty from an ensemble model, which was trained on time delay data with no jitter noise, for trajectories generated using varying levels of time jitter noise. 
We can see in this case the epistemic uncertainty again exhibits more sensitivity to data drift for the full range of values of the detuning.
This can be useful as a way of detecting imperfections or miscalibration in the experiential setup (in this case detector defectiveness) used to generate the time delay results given to the model at inference time.

\subsection{Posterior approximation}
\begin{figure}
    \centering
    \includegraphics[width=1.025\columnwidth]{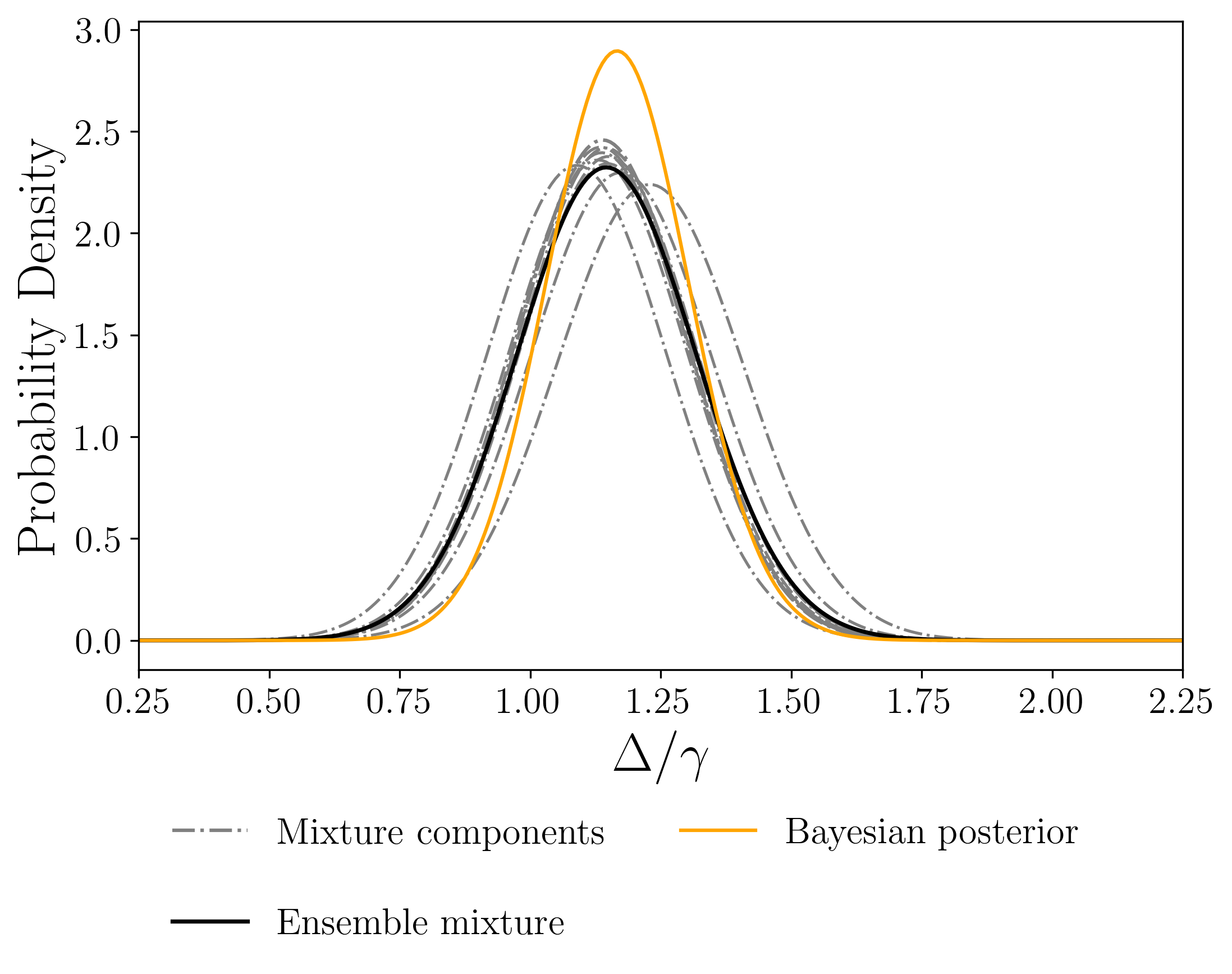}
    \caption{Posterior distribution computed with a uniform prior over $\Delta \in [0, 5\gamma]$ and mixture distribution learned by the ensemble for an example trajectory generated with $\Delta\approx1$ and $\Omega=1$. }
    \label{fig:dist_example}
\end{figure}

In addition to providing uncertainty estimates another benefit provided by our ensemble approach, which is similarly unavailable for currently existing single network approaches, is the ability to provide a smooth probability distribution over possible parameter values similar to the Bayesian inference approach. 
Naturally one may ask, given the uniform prior used in the Bayesian analysis and the universal approximation property of Gaussian mixtures, if we can expect the ensembles output distribution to have good fidelity with the Bayesian posterior. 
We provide an example of the Bayesian posterior and the ensemble mixture distribution for a single trajectory in Fig. \ref{fig:dist_example} which suggests that this is indeed possible.
To examine this further we compute the average fidelity between the Bayesian posterior distribution and the distribution learned by the ensemble over 100 random trajectories for each unique value of $\Delta$ in our test set.
As done in Ref.~\cite{clark2025efficient} we compute the fidelity between the analytic posterior and the mixture distribution learned by the ensemble as the Bhattacharyya coefficient 
\begin{align}
    F(p_A(\theta), p_B(\theta)) = \int_{\Theta} \sqrt{p_A(\theta)p_B(\theta)}d\theta
\end{align}
whose range is $[0,1]$.
\begin{figure}
    \centering
    \includegraphics[width=1.025\columnwidth]{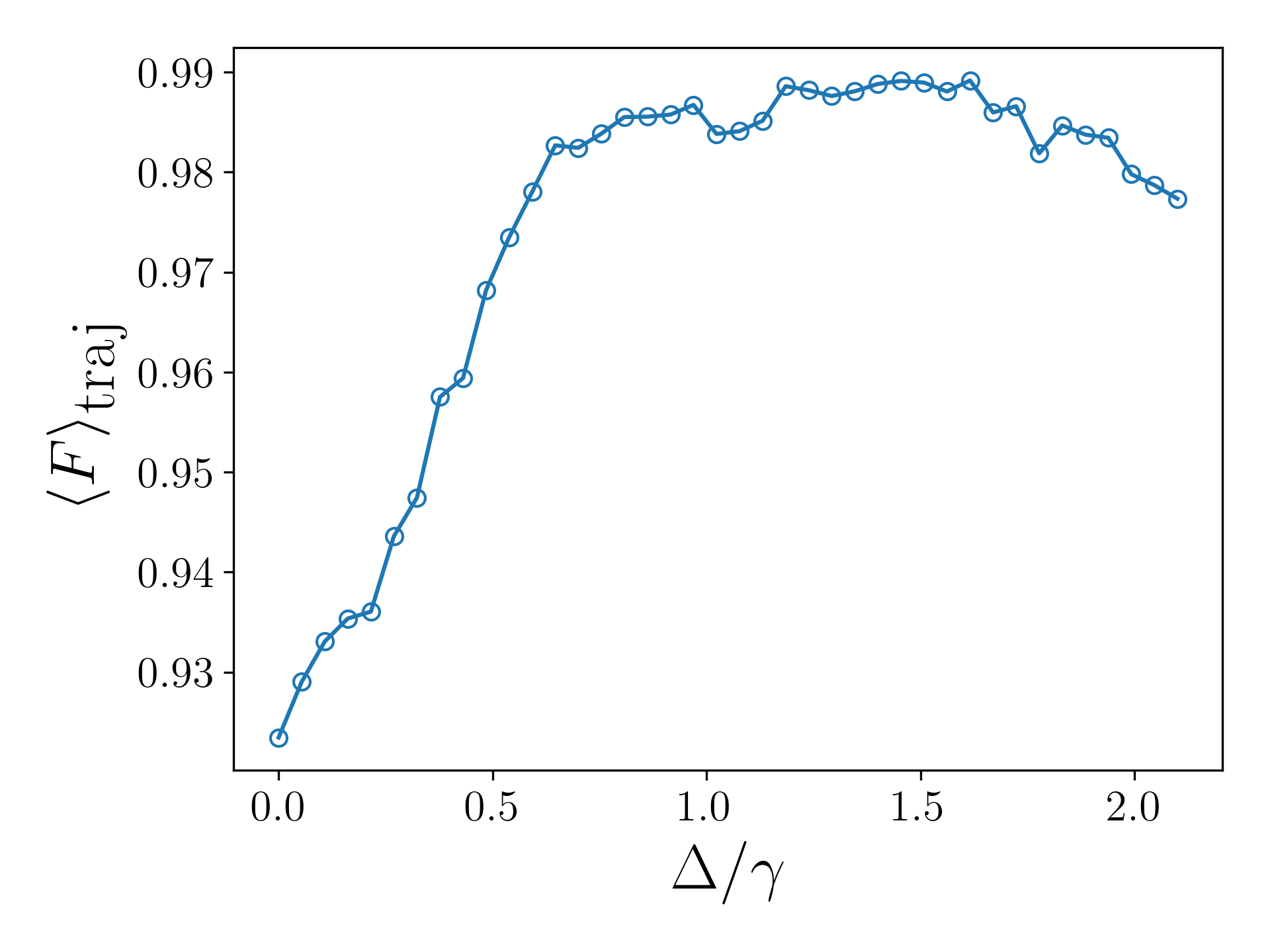}
    \caption{Average Bhattacharyya fidelity between Bayesian posterior and ensemble Gaussian mixture distribution over 100 random trajectories per $\Delta$ value. }
    \label{fig:ensemble_fid}
\end{figure}

\begin{figure*}
    \centering
    \includegraphics[width=1.025\textwidth]{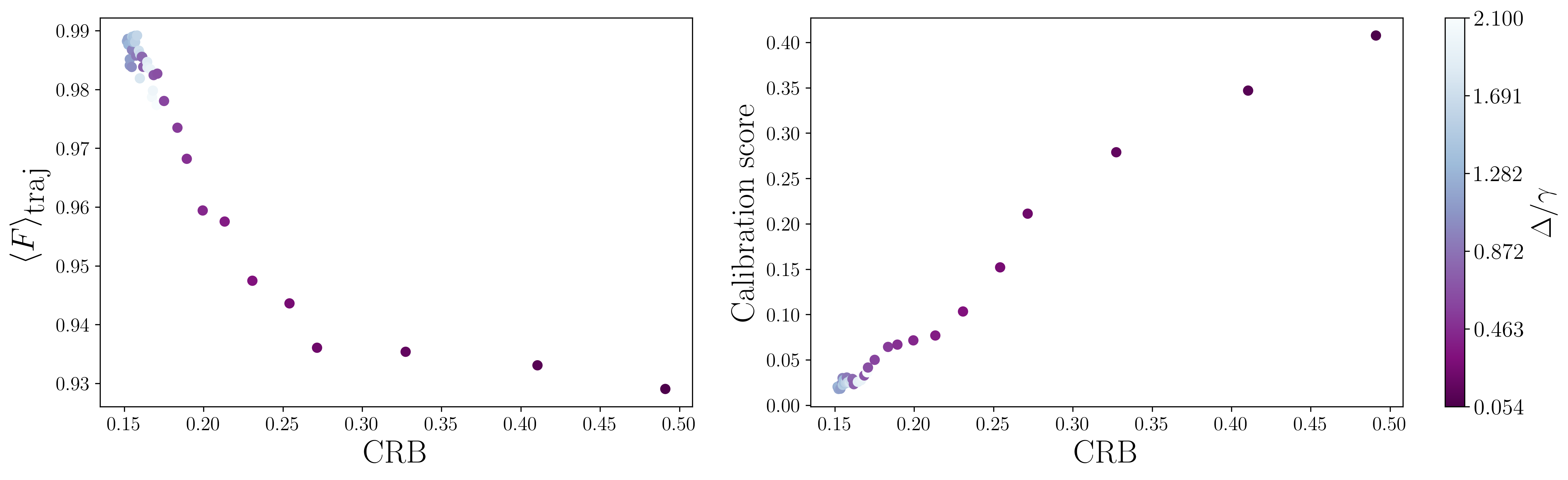}
    \caption{Average Bhattacharyya fidelity between Bayesian posterior and ensemble Gaussian mixture distribution and average calibration score as a function of the unbiased CRB. Note we do not plot the CRB for $\Delta=0$ as it diverges for this value.}
    \label{fig:ece_fid_crb}
\end{figure*}

The results are shown in Fig. \ref{fig:ensemble_fid}. 
We can see that for the majority of $\Delta$ values the distribution learned by the ensemble has strong (near unity) overlap with the Bayesian posterior. 
The fidelity degrades slightly near resonance values $\Delta \approx 0$ as the uniform prior leads to an abrupt vanishing of the posterior density for values of $\Delta<0$ leading to a non-smoothly varying distribution near these values unlike the one always produced by the ensemble. 
Additionally, these values being close to the limit of the range of $\Delta$ values, $[0, 5\gamma]$, used in the training data of the ensemble likely exacerbates the discrepancies.

\subsection{Calibration assessment}

Here we examine the calibration performance of the ensemble by computing the expected calibration error (ECE). 
In particular, we use an adaptation of the ECE metric for regression tasks called the calibration score~\cite{wibbeke2026evaluating}.
Given a set of $m$ confidence thresholds $p_j$ such that $0 \leq p_1 \leq \cdots \leq p_m \leq 1$ we compute the empirical frequency
\begin{equation}
    \hat{p}_j = \frac{\left|\{y_i \mid \hat{F}_i(y_i) \leq p_j,\ i = 1,\dots,N\}\right|}{N},
\end{equation}
where $N$ is the size of the dataset, $y_i$ is the ground truth label corresponding to data point $x_i$, $\hat{F}_i$ denotes the cumulative distribution function (CDF) of the model's predicted distribution for data point $x_i$. 
In our case $\hat{F}_i(y_i)=\Phi\left(\frac{y_i-\mu(x_i)}{\sigma(x_i)}\right)$ is the Gaussian CDF.
The normalized calibration score is then defined as
\begin{equation}
    \mathrm{CalS}(\hat{F},y) = \sqrt{\frac{1}{m}\sum_{j=1}^{m} \left(p_j - \hat{p}_j\right)^2}.
\end{equation}
Intuitively, the calibration score measures how closely the empirical coverage $\hat{p}_j$ tracks the nominal confidence level $p_j$ across all thresholds and therefore a well-calibrated model should have $\hat{p}_j \approx p_j$ for every $j$.

We plot the fidelity along with the calibration score, for $m=10$ evenly spaced confidence thresholds, against the unbiased CRB in Fig. \ref{fig:ece_fid_crb}. 
We do this by first computing the calibration score for each unique value of the detuning and plotting it against it's corresponding CRB. 
For a large majority of values of the CRB we see that the calibration score is low indicating that the model is well calibrated. 
However, we see that both the fidelity and calibration score degrade as the CRB increases.
From the heatmap we can see that these higher values of the CRB correspond to detuning values close to resonance.
This follows the same trend as the high RMSE attained in previous sections for the Bayesian estimator and the neural network models in this detuning regime. 
From prior findings in the literature~\cite{kiilerich2014estimation} it is known that the Fisher information for photon counting, as implemented here, vanishes as $\Delta \rightarrow 0$ leading to the CRB quickly diverging. 
This likely explains the poor performance of all estimators as it means the photon counting measurement can not distinguish values of $\Delta$ in this neighborhood of the parameter space.

\subsection{Multi-parameter estimation}
\begin{figure*}
    \centering
    \includegraphics[width=1.025\textwidth]{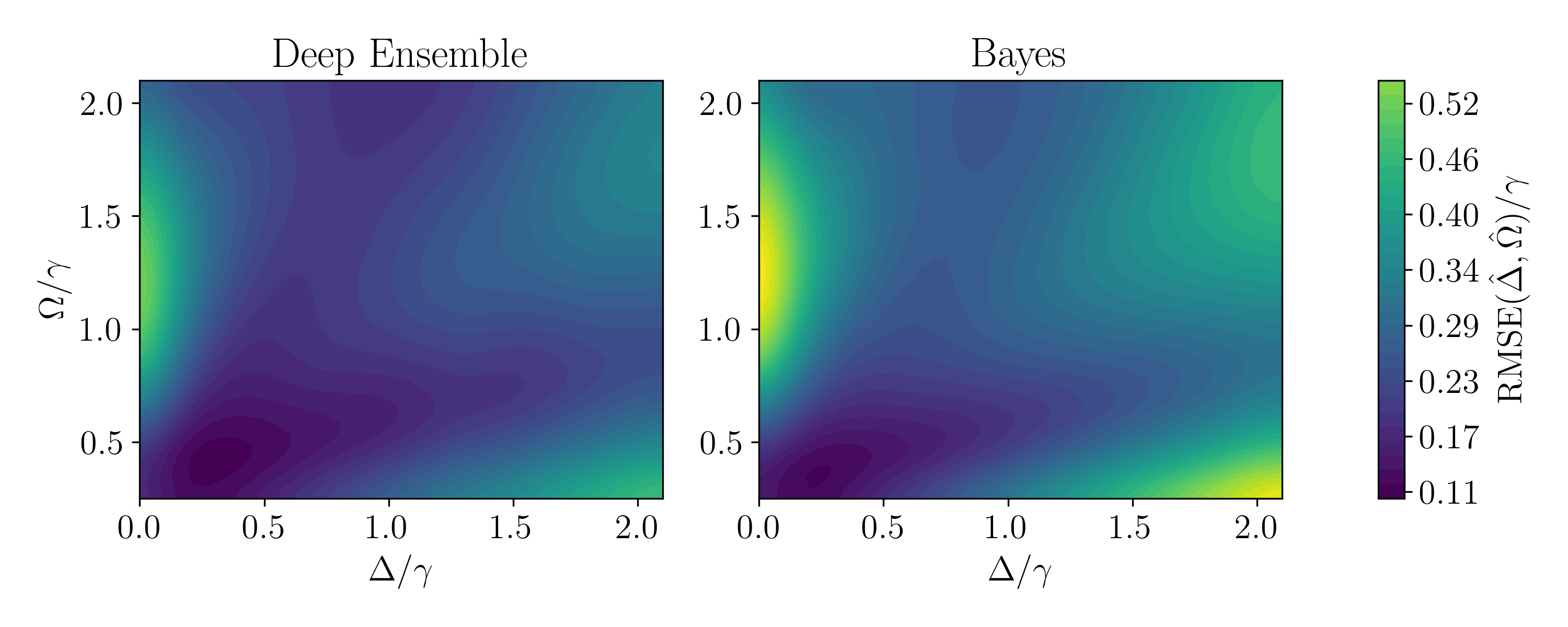}
    \caption{Contour plots of average RMSE achieved by the deep ensemble and Bayesian inference via nested sampling on test set.}
    \label{fig:2d_rmse}
\end{figure*}

Here we demonstrate our approach applied to the simultaneous estimation of the Rabi frequency $\Omega$ and detuning $\Delta$.
We train an ensemble of $M=10$ networks on $16 \times 10^4$ trajectories of length $N=48$ which were generated using random values of $\Delta \in [0, 3\gamma]$ and $\Omega \in [0.25\gamma, 5\gamma]$ drawn uniformly from their respective intervals.
We compare our ensembles performance to that of standard Bayesian inference performed using nested sampling~\cite{buchner2023nested, feroz2008multimodal}. 
To avoid the prohibitively long inference time on our computational resources associated with the two-dimensional Bayesian inference routine we evaluate the trained deep ensemble on the test set of trajectories from Ref.~\cite{rinaldi2024parameter} whose precomputed Bayesian point estimates for the test set trajectories are made available at~\cite{sanchez_munoz_2023_8305509}.
The test set contains $16 \times 10^6$ trajectories to ensure convergence of the computed RMSEs. 
A uniform prior used over the interval from which the training set was generated from to ensure fair comparison with the ensemble.
Fig. \ref{fig:2d_rmse} shows the RMSE achieved by both the deep ensemble and Bayesian inference. 
We can see that both approaches perform comparably with the deep ensemble often achieving a lower RMSE particularly for parameter values close to the boundaries of the parameter space under consideration.
This shows that not only can the deep ensemble exceed the inference speed of likelihood-based Bayesian inference but that it can also achieve better performance in terms of estimation accuracy.

\subsection{Comparison to Approximate Bayesian Computation}\label{abc_comp}

Here we will compare the performance of our likelihood-free parameter inference method with the state-of-the-art likelihood-free method presented in Ref. \cite{clark2025efficient} i.e. approximate Bayesian computation (ABC).
For this comparison we consider the more complex optomechanical system studied in the same work. 
\subsubsection{Quantum system}
The system consists of an optical cavity which is driven by an external laser with one of the mirrors being free to oscillate which constitutes the mechanical mode of the system~\cite{clark2022exploiting, clark2025efficient}. 
The Hamiltonian of the closed system is given by
\begin{align}{\label{ham_om}}
\hat{H} = -\Delta\hat{a}^\dag \hat{a} + \omega_M \hat{b}^\dag \hat{b} + \frac{\Omega}{2}(\hat{a} + \hat{a}^\dag) + g \hat{a}^\dag \hat{a} (\hat{b} + \hat{b}^\dag)
\end{align}
where $\hat{a}$ and $\hat{b}$ are the annihilation operators of the optical and mechanical modes respectively, $\Omega$ is the laser Rabi frequency, $g$ is the optomechanical coupling strength, and $\omega_M$ is the mechanical frequency.
The mechanical mode interacts with a thermal bath by both emitting and absorbing phonons while the optical mode leaks photons which are continuously detected. 
The master equation for the open system is given by 
\begin{align}{\label{master_eq_om}}
\frac{\partial }{\partial t}\hat{\rho}(t) = &-i[\hat{H},\hat{\rho}(t)] +  
\kappa\left(\hat{a}\hat{\rho}(t)\hat{a}^\dag - \frac{1}{2}\{\hat{a}^\dag\hat{a}, \hat{\rho}(t)\}\right) \\
&+\gamma(\bar{m}+1)\left(\hat{b}\hat{\rho}(t)\hat{b}^\dag - \frac{1}{2}\{\hat{b}^\dag\hat{b}, \hat{\rho}(t)\}\right)\\
&+\gamma\bar{m}\left(\hat{b}^\dag\hat{\rho}(t)\hat{b} - \frac{1}{2}\{\hat{b}\hat{b}^\dag, \hat{\rho}(t)\}\right)
\end{align}
where $\bar{m}=(\exp{(\omega_M/k_BT)-1})^{-1}$ is the mean phonon number of the bath, $T$ is the temperature of the bath, $\kappa$ is the decay rate of the optical mode and $\gamma$ is the decay rate of the mechanical mode.

Specifically, we consider the nonlinear regime of the system which arises when the coupling strength between the optical and mechanical modes is strong~\cite{clark2025efficient} and therefore the analytic Gaussian formalism is no longer sufficient to describe the system dynamics~\cite{clark2022exploiting}.
In this case no analytic solution to the master equation is available~\cite{clark2025efficient}, which precludes the derivation of a likelihood function, thus necessitating the use of likelihood-free methods such as ABC.
Notably one can set the detuning value as $\Delta_n=-ng^2/\omega_M$, where $n \in \mathbb{Z}^+$, so that $n$-photon emissions become favored by the system resulting in photon bunching in the $n\geq2$ regime~\cite{clark2022exploiting, clark2025efficient}.
Note that the data used here has $g=4$ and $\omega_M=4\sqrt{2}$~\cite{clark2025efficient}.

\subsubsection{Predictive accuracy benchmark}
We train our deep ensemble to again minimize the Gaussian negative log likelihood loss.
For the training data we use the data library from Ref. \cite{clark2025efficient}.
The data contains time delay trajectories of $80$ photon detections for 100 evenly spaced values of the detuning $\Delta$ across the range $[-10, 0]$ with $4,000$ trajectories for each unique $\Delta$. 
We randomly split the library using an 80/20 split between training and testing data and stratify by $\Delta$ value to ensure each set is balanced and that we use a uniform prior for the ABC algorithm as done in~\cite{clark2025efficient}.
Due to the nature of the dynamics of the system in the nonlinear regime, i.e. the system can generally not be approximated as a coherent state and assumed to reset to a consistent state with each photon detection like in the case of the TLS, the assumption of independence and identical distribution of the time delays in a given trajectory no longer holds.
While these additional time delay correlations allow for potential improvement in the ability to perform sensing tasks~\cite{clark2022exploiting} they also lead to the smoothed histogram neural network layer used previously no longer being optimal.

Two tools that are more appropriate in this setting are recurrent neural network layers, such as a long-short term memory layer, or a one-dimensional convolutional layer.
We use the latter as it is generally simpler to implement and more amenable to parallelization than recurrent architectures~\cite{vaswani2017attention}. 
The specific architecture of the deep ensemble networks are given in Appendix \ref{conv_net_appendix}.
We use an ensemble of $M=10$ networks with $\beta_1=0.87$, $\beta_2=0.93$, a learning rate of $10^{-3}$ for the Adam optimizer, and a batch size of $250$ over 50 epochs.

When benchmarking the ABC method we treat the training data as the data library. 
While the pseudocode for the ABC algorithm is given in Ref.~\cite{clark2025efficient} we give a more detailed implementation in Algorithm \ref{alg:abc} for completeness. 
\begin{figure*}[t]
\centering
\begin{minipage}{0.95\textwidth}
\begin{algorithm}[H]
\caption{Approximate Bayesian computation algorithm}
\label{alg:abc}
\begin{algorithmic}[1]
\Require Library of trajectory-parameter pairs $\mathcal{D}$, number of times to sample from the library $\nu$, acceptance threshold for sample $\epsilon$, summary statistic function $S(\cdot)$, discrepancy measure function $\delta(\cdot)$, observed trajectory to perform parameter inference on $D$
\State Initialize empty histogram $H$ of $\theta$'s 
\State $i \leftarrow 0$
\While{$i < \nu$}
\State Sample, without replacement, one trajectory-parameter pair $(D',\theta)$ from $\mathcal{D}$ 
\If{$\delta(S((D',\theta)), S(D)) \leq \epsilon$}
\State  $H[\theta] \leftarrow H[\theta]+1$  \Comment{Add one count to histogram bin corresponding to value $\theta$}
\EndIf
\State $i \leftarrow i + 1$
\EndWhile
\State Normalize $H$ \\
\Return $H$ \Comment{Return the normalized histogram which serves as an approximation of the Bayesian posterior}
\end{algorithmic}
\end{algorithm}
\end{minipage}
\end{figure*}
We draw $\nu=10^5$ samples and set $\epsilon=9$ as this was found to be more than enough for the method to converge for estimation of the same parameter and data in~\cite{clark2025efficient}.
We also use the same histogram based summary statistic and $L_2$ discrepancy measure from~\cite{clark2025efficient} and similarly take the mean of the ABC posterior as our estimator of the parameter.
\begin{figure}
    \centering
    \includegraphics[width=1.025\columnwidth]{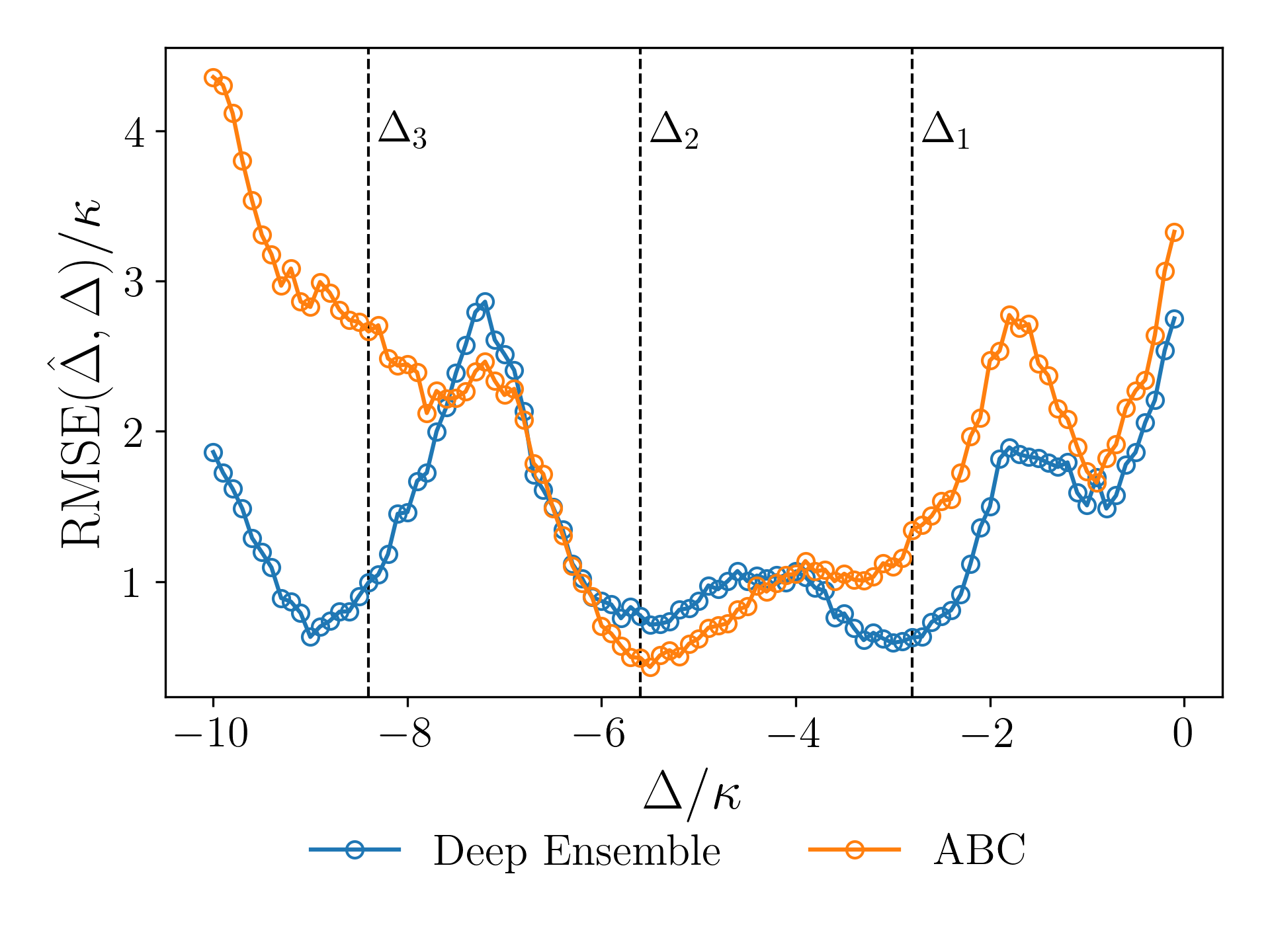}
    \caption{Performance of ABC and deep ensemble on the test set for the nonlinear optomechanical system.}
    \label{fig:abc_ensemble_rmse}
\end{figure}
Fig. \ref{fig:abc_ensemble_rmse} shows the performance of the two methods on our test set. 
We can see that for most values of $\Delta$ the deep ensemble achieves lower error than ABC and achieves competitive accuracy in all other cases. 
Interestingly, for the $\Delta_{n=3}$ regime we see that the ensemble greatly outperforms the ABC method suggesting it makes better use of the information encoded in the three-photon correlations present in the trajectory data.
Note also that the inference time for the ABC method is also about two orders of magnitude larger as we shall see in the next section.

\section{Computational Efficiency}

\begin{figure}
    \centering
    \begin{subfigure}[b]{0.475\textwidth}
    \centering
    \includegraphics[width=1.025\columnwidth]{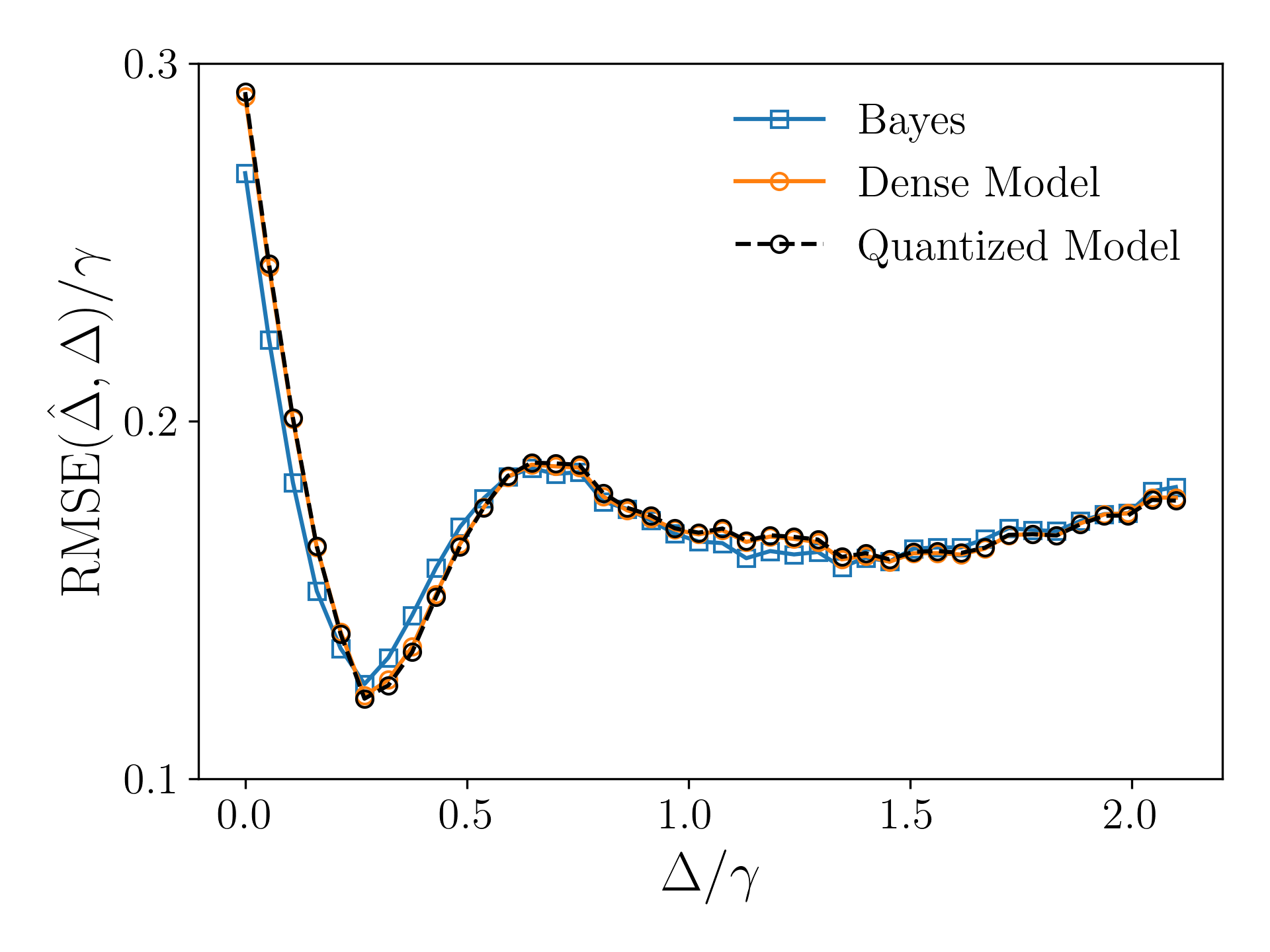}
    \caption{Average RMSE of non-quantized deep ensemble and quantized deep ensemble.}
    \label{fig:quant_rmse}
    \end{subfigure}
    \begin{subfigure}[b]{0.475\textwidth}
    \centering
    \includegraphics[width=1.025\columnwidth]{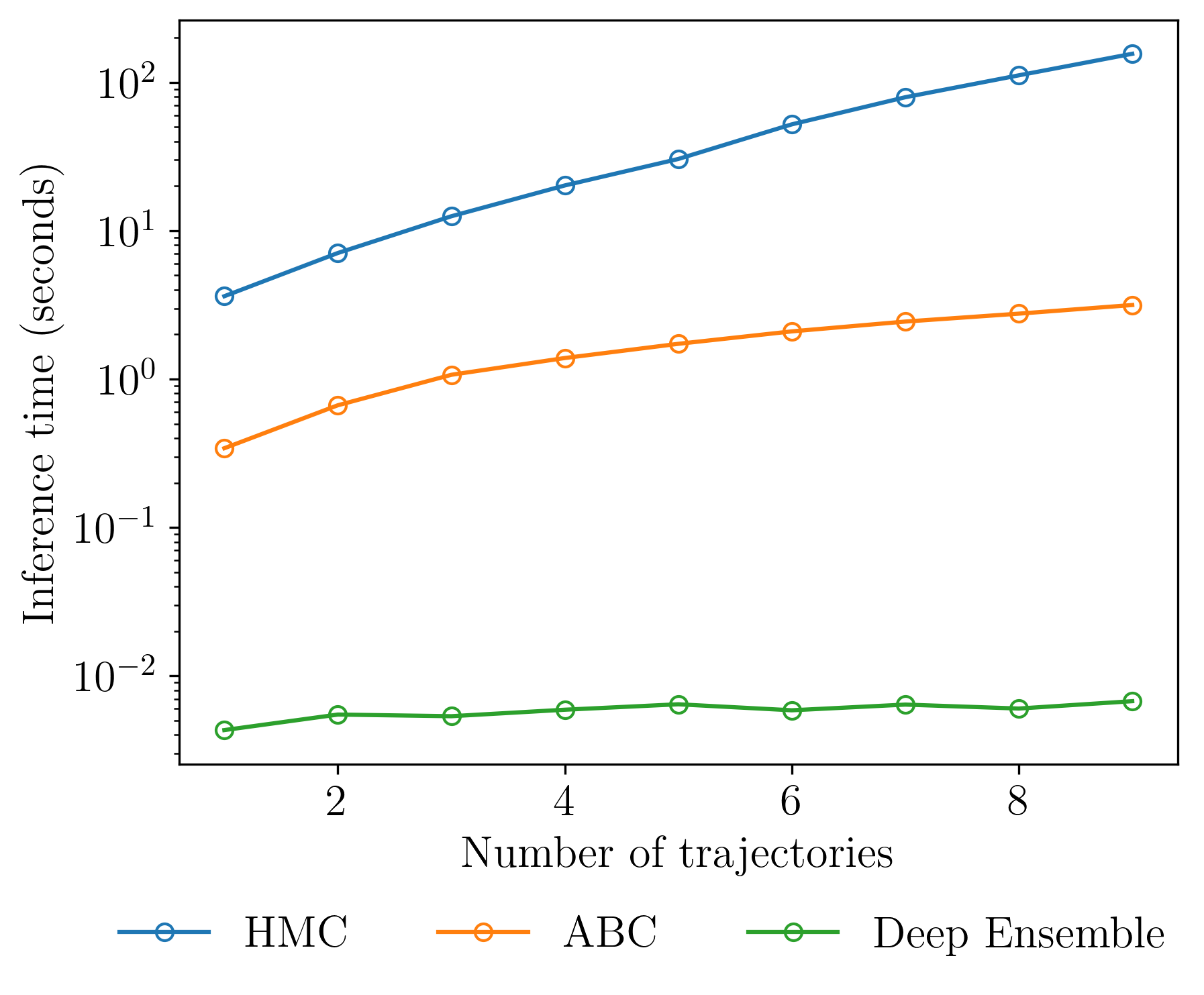}
    \caption{Average inference times over 10 runs for HMC based and ABC based Bayesian inference and quantized deep ensemble for various numbers of trajectories.}
    \label{fig:inference_time}
    \end{subfigure}
    \caption{Predictive accuracy and inference time of Bayesian inference and  quantized deep ensemble.}
\end{figure}

Here we more closely examine the advantage provided by NNs in terms of their computational efficiency with respect to  inference time and memory overhead particularly when compressed for deployment on edge devices such as FPGAs. 
The deep ensembles discussed thus far have a memory size of roughly 1.1 megabtyes. This can be further compressed via the use of model quantization which converts the 32-bit floating point weights of the model to 8-bit integers. 
This reduces the size of our models to roughly 8.8 kilobytes. 
This contrasts with a method such as ABC which requires the storage of the data library during inference time and thus has a space complexity which is linear in the size of the dataset. 
This is valuable in cases where one wishes to deploy these models to edge devices with limited memory to perform real time inference and does not lead to any noticeable deterioration of model performance as can be seen in Fig. \ref{fig:quant_rmse}.
Other memory optimization methods, such as weight pruning, can be performed if further memory efficiency is required. 

We also compare inference times by measuring the wall clock time of a trained quantized deep ensemble, the ABC method and an optimized likelihood-based Hamiltonian Monte Carlo (HMC) Bayesian inference routine for estimating $\Delta$ given delay measurements from different numbers of trajectories. 
We use the state-of-the-art No U-Turn Sampler (NUTS) provided by the \texttt{PyMC} framework as our specific HMC method \cite{hoffman2014no, abril2023pymc}. For inference with NUTS we use 4 chains and draw 1,000 posterior samples per chain which we parallelize over 4 cores. 
For inference with ABC we use the same implementation of the algorithm used in the previous section which leverages \texttt{NumPy} array vectorization~\cite{harris2020array} when possible to ensure fair comparison. 
To ensure fairness in the comparison we perform inference only on the CPU with all methods. However, we note that deep ensembles can be made even more scalable when performing inference on a GPU especially when utilizing convolutional layers like in the case of the optomechanical system.
The average inference time for the three methods over 20 different runs of the inference on an AMD Ryzen 9 3900X 12-Core Processor are shown in Fig. \ref{fig:inference_time}. 
We can see that the highly parallelizable nature of neural network inference, i.e. batched inference, allows for orders of magnitude faster inference speed over the NUTS and ABC method despite the need to perform a forward pass for each of the 10 NNs in the deep ensemble. 
The ability to parallelize inference with a neural network, due to the fact that the network is carrying out simple linear algebraic operations, contrasts with the inherently sequential nature of MCMC based Bayesian inference, due to the construction of the Markov chains, and ABC based Bayesian inference due to the rejection sampling routine. 
This shows that deep ensembles are better equipped for real time quantum parameter estimation tasks while still retaining the uncertainty quantification capabilities of Bayesian inference.  

\section{Conclusion}
We have proposed a method for using NNs to perform parameter estimation which provides uncertainty estimates in addition to point estimates of the parameters. 
We showed that these models provide good parameter estimates despite being optimized for both predictive accuracy and uncertainty estimation. We also show that they provide well calibrated uncertainty estimates and good approximations of the Bayesian posterior distribution.
We also demonstrated our scheme is applicable to more complex systems in which the measurement results are not independent and identically distributed, unlike the case of the TLS studied in much of the present work, where recurrent or convolutional NNs become more appropriate to use over the permutation-invariant architectures discussed initially. 
Furthermore, our method can, in principle, be generalized to $n$-dimensional estimation problems by only adding an additional $O(n)$ neurons in the output layer of the network as opposed to classification based approaches which require $O(m^n)$ where $m$ is the resolution of the discretized parameter space~\cite{nolan2021machine, cimini2023deep}.
We have demonstrated this for the $n=2$ case in the present work.
We leave higher dimensional extensions for future work along with the exploration of other NN architectures, such as deep sets, that possess permutation invariance and the ability to naturally handle inputs of varying length \cite{zaheer2017deep}. 

In this work we performed hyperparameter tuning on a single model and used the optimal hyperparameters from that search to train the deep ensemble due to time constraints. An interesting question to answer would be if hyperparameter tuning on the ensemble level, while being more computationally expensive, could deliver even better results. 
We also only considered ensembles in which all models have the same architecture but this is by no means a necessity. 
Further exploration of how using ensembles of networks with different architectures, e.g. different activation functions and numbers of hidden layers, would impact ensemble performance on the parameter estimation task would be useful, in particular with respect to how this affects the uncertainty estimates as this approach may better capture epistemic (model) uncertainty.
Finally, since photon counting is not necessarily the most optimal measurement in terms of the quantum Fisher information (QFI)~\cite{rinaldi2024parameter} it would be interesting to explore how the performance and data efficiency of these models changes when the QFI of the continuous measurement is optimized~\cite{chinni2026optimal}.

\textit{Note}: Some time after our manuscript appeared (arXiv:2509.10756), a work was posted (arXiv:2602.19792) which also introduced methods for performing quantum parameter estimation using neural networks~\cite{molenda2026unlocking}. 
This approach utilizes single networks, however, which could make quantifying epistemic uncertainty more challenging. 
It also utilizes the Gaussian negative log likelihood loss as done here but learns the covariances of parameters in addition to the means and variances which may be useful. 

\begin{acknowledgments}
We thank Carlos S\'anchez Mu\~noz, Tyree Giles, Derek Everett, Olivier Pfister and Aaron Reding for useful discussions. 
\end{acknowledgments}

\appendix

\section{Bayesian Inference}{\label{appendix:A}}
The objective of Bayesian inference is to compute the posterior probability distribution $P(\theta|D)$ where $\theta$ is the parameter being estimated and $D$ represents the observed data. 
From Bayes' rule the posterior is given by

\begin{align}
    P(\theta|D) = \frac{P(D|\theta)P(\theta)}{P(D)}
\end{align}
where $P(D|\theta)$ is the likelihood, $P(\theta)$ is the prior distribution over parameter values which can be chosen to incorporate our prior knowledge about the parameter, and $P(D)$ is called the evidence or the marginal likelihood \cite{lambert2018student, gelman1995bayesian}. 
The Bayesian estimator, $\hat{\theta}$, of $\theta$ is chosen to be either the mean, median or mode of the posterior with the posterior mean being the most common choice \cite{devore2021modern}. In our work we use the posterior mean $\mathbb{E}[P(\theta|D)]$.
The delay time probability distribution for the TLS, $w(\tau;\Delta, \Omega, \gamma)$, which gives the probability density for a particular delay time $\tau$ is given by \cite{rinaldi2024parameter}:

\begin{multline}
w(\tau;\Delta, \Omega, \gamma) = \frac{8\gamma\Omega^2}{R} e^{-\gamma\tau/2} 
\\ 
\times  \left( \sum_{\zeta=-1,1}\zeta\cosh{\left(\frac{\tau\sqrt{\gamma^2-4(\Delta^2+4\Omega^2)+\zeta R}}{2\sqrt{2}}\right)} \right) 
\end{multline}
where $R=\sqrt{[\gamma^2+4(\Delta^2+4\Omega^2)]^2-64\gamma^2\Omega^2}$. Using this the likelihood for a given set of time delay measurements $D=[\tau_1,...,\tau_N]$ is given by

\begin{align}
    P(D|\theta) = \prod_{i=1}^N w(\tau_i;\Delta, \Omega, \gamma).
\end{align}
For our prior we use a uniform distribution with support on the range the training data was genereated from for all experiments with the TLS.

The analytic expression of the posterior distribution is in general intractable to compute due to the denominator in Bayes' rule except for special cases such as conjugate prior-likelihood pairs. 
Due to this one must usually resort to simply drawing samples from the posterior distribution and computing summary statistics from these samples to perform inference. Markov chain Monte Carlo methods which draw (dependent) samples from our posterior distributions are the most common methods used in Bayesian inference as they completely circumvent the need to compute the denominator in Bayes rule \cite{lambert2018student}. An alternative approach which is more common in astrophysical studies is to use nested sampling \cite{buchner2023nested, feroz2008multimodal}. 
For the results presented in Section \ref{ml_exp} we resort to numerically computing the posterior distribution by evaluating the denominator over a fine grid of 500 points on the support of the prior. Note that we do this as it is more exact for the case of a one-dimensional estimation problem but quickly becomes unwieldy for multi-dimensional problems.   

\section{Hyperparameter Tuning Details}{\label{appendix:B}}

Hyperparameter tuning is an integral part of the ML training process particularly the \textit{model selection} stage \cite{james2023statistical, goodfellow2016deep}. If the number of hyperparameters is small and the possible values each can take on are as well then a simple grid search can be used to find the optimal set of hyperparameters. However this is impractical for most modern deep learning models which can have many hyperparameters many of which are continuous in nature. Therefore much work has been done on hyperparameter tuning methods which are both fast and converge on a good set of hyperparameters. 
Methods based on Bayesian optimization such as the tree-structured Parzen estimator are the current state-of-the-art for hyperparameter tuning \cite{akiba2019optuna}. Table \ref{tab:hyperparam_table} shows the hyperparameters that were tuned during the model selection process as well as the search spaces for each hyperparameter. 

\begin{table}[ht]
\centering
\begin{tabular}{|c|c|}
\hline
Hyperparameter & Search Space \\
\hline
Number of histogram bins & $(200, 710)$ \\
\hline
Loss function & \{RMSE, MSLE\} \\
\hline
Learning rate & $(10^{-5}, 5 \times 10^{-3})$ \\
\hline
$\beta_1$ of Adam optimizer & $(0.8, 0.999)$ \\
\hline
$\beta_2$ of Adam optimizer & $(0.8, 0.999)$ \\
\hline
Number of epochs & $(50, 500)$ \\
\hline
Batch size & $\{2^6, 2^7, 2^8, 2^9, 2^{10}, 2^{11}\}$ \\
\hline
Dropout probability & $(0, 0.2)$ \\
\hline
Early stopping patience & $(4, 10)$ \\
\hline
\end{tabular}
\caption{Hyperparameters tuned during model selection with associated search spaces. Values in parenthesis indicate the lower and upper range of a search space whereas values in braces indicate the full discrete set of parameter values. }
\label{tab:hyperparam_table}
\end{table}

\section{Convolutional neural network architecture}\label{conv_net_appendix}
Here we provide the specific architecture of the convolutional neural network used in Section \ref{abc_comp} in Table \ref{conv_net}.

\begin{table}[ht]
\centering
\caption{Neural network architecture. The notation $M \rightarrow N$ denotes a mapping from $M$ features to $N$ features.}
\label{conv_net}

\begin{tabularx}{0.5\textwidth}{llX}
\hline
\textbf{Module} & \textbf{Configuration} \\
\hline

\multicolumn{3}{l}{\textit{Convolutional Layers}} \\
  Conv1d & $1 \rightarrow 100$, kernel: 6, stride: 1, padding: 1 \\
 BatchNorm1d & 100 features \\
 ReLU & activation \\
 Conv1d & $100 \rightarrow 200$, kernel: 6, stride: 1, padding: 1 \\
 BatchNorm1d & 200 features \\
 ReLU & activation \\

\hline
\multicolumn{3}{l}{\textit{Dense Layers}} \\
 Linear & $200 \rightarrow 128$ \\
 BatchNorm1d & 128 features \\
 ReLU & activation \\
 Linear & $128 \rightarrow 64$ \\
 BatchNorm1d & 64 features \\
 ReLU & activation \\
 Linear & $64 \rightarrow 32$ \\
 BatchNorm1d & 32 features \\
 ReLU & activation \\

\hline
\multicolumn{3}{l}{\textit{Output Layer}} \\
 Linear & $32 \rightarrow 2$ \\
\hline
\end{tabularx}
\end{table}

\bibliography{Anteneh}

@article{wibbeke2026evaluating,
  title={Evaluating the quality of the quantified uncertainty for (Re) calibration of data-driven regression models},
  author={Wibbeke, Jelke and Sch{\"o}nfisch, Nico and Rohjans, Sebastian and Rauh, Andreas},
  journal={International Journal of Approximate Reasoning},
  pages={109685},
  year={2026},
  publisher={Elsevier}
}

@article{molenda2026unlocking,
  title={Unlocking photodetection for quantum sensing with Bayesian likelihood-free methods and deep learning},
  author={Molenda, Mateusz and Clark, Lewis A and P{\l}odzie{\'n}, Marcin and Kolodynski, Jan},
  journal={arXiv preprint arXiv:2602.19792},
  year={2026}
}

@article{gammelmark2014fisher,
  title={Fisher information and the quantum Cram{\'e}r-Rao sensitivity limit of continuous measurements},
  author={Gammelmark, S{\o}ren and M{\o}lmer, Klaus},
  journal={Physical review letters},
  volume={112},
  number={17},
  pages={170401},
  year={2014},
  publisher={APS}
}

@article{barbieri2022optical,
  title={Optical quantum metrology},
  author={Barbieri, Marco},
  journal={Prx Quantum},
  volume={3},
  number={1},
  pages={010202},
  year={2022},
  publisher={APS}
}

@article{chinni2026optimal,
  title={Optimal waveforms for dipole moment estimation with coherent states},
  author={Chinni, Karthik and Quesada, Nicol{\'a}s},
  journal={New Journal of Physics},
  volume={28},
  number={3},
  pages={034507},
  year={2026},
  publisher={IOP Publishing}
}

@article{harris2020array,
  title={Array programming with NumPy},
  author={Harris, Charles R and Millman, K Jarrod and Van Der Walt, St{\'e}fan J and Gommers, Ralf and Virtanen, Pauli and Cournapeau, David and Wieser, Eric and Taylor, Julian and Berg, Sebastian and Smith, Nathaniel J and others},
  journal={nature},
  volume={585},
  number={7825},
  pages={357--362},
  year={2020},
  publisher={Nature Publishing Group UK London}
}

@article{clark2022exploiting,
  title={Exploiting non-linear effects in optomechanical sensors continuous with photon-counting},
  author={Clark, Lewis A and Markowicz, Bartosz and Ko{\l}ody{\'n}ski, Jan},
  journal={Quantum},
  volume={6},
  pages={812},
  year={2022},
  publisher={Association for the F{\"o}demotion of Open Access Publishing in Quantum Sciences}
}

@article{vaswani2017attention,
  title={Attention is all you need},
  author={Vaswani, Ashish and Shazeer, Noam and Parmar, Niki and Uszkoreit, Jakob and Jones, Llion and Gomez, Aidan N and Kaiser, {\L}ukasz and Polosukhin, Illia},
  journal={Advances in neural information processing systems},
  volume={30},
  year={2017}
}

@phdthesis{thuy2025learning,
  title={Learning Analytics with Neural Networks: Addressing Open Challenges Through Uncertainty Quantification and Natural Language Processing},
  author={Thuy, Arthur},
  year={2025},
  school={Ghent University}
}

@article{thuy2024explainability,
  title={Explainability through uncertainty: Trustworthy decision-making with neural networks},
  author={Thuy, Arthur and Benoit, Dries F},
  journal={European Journal of Operational Research},
  volume={317},
  number={2},
  pages={330--340},
  year={2024},
  publisher={Elsevier}
}

@article{thuy2024fast,
  title={Fast and reliable uncertainty quantification with neural network ensembles for industrial image classification},
  author={Thuy, Arthur and Benoit, Dries F},
  journal={Annals of Operations Research},
  pages={1--27},
  year={2024},
  publisher={Springer}
}

@article{lee2015m,
  title={Why m heads are better than one: Training a diverse ensemble of deep networks},
  author={Lee, Stefan and Purushwalkam, Senthil and Cogswell, Michael and Crandall, David and Batra, Dhruv},
  journal={arXiv preprint arXiv:1511.06314},
  year={2015}
}

@book{goodfellow2016deep,
  title={Deep learning},
  author={Goodfellow, Ian and Bengio, Yoshua and Courville, Aaron and Bengio, Yoshua},
  volume={1},
  number={2},
  year={2016},
  publisher={MIT press Cambridge}
}

@article{abril2023pymc,
  title={PyMC: a modern, and comprehensive probabilistic programming framework in Python},
  author={Abril-Pla, Oriol and Andreani, Virgile and Carroll, Colin and Dong, Larry and Fonnesbeck, Christopher J and Kochurov, Maxim and Kumar, Ravin and Lao, Junpeng and Luhmann, Christian C and Martin, Osvaldo A and others},
  journal={PeerJ Computer Science},
  volume={9},
  pages={e1516},
  year={2023},
  publisher={PeerJ Inc.}
}

@inproceedings{akiba2019optuna,
  title={Optuna: A next-generation hyperparameter optimization framework},
  author={Akiba, Takuya and Sano, Shotaro and Yanase, Toshihiko and Ohta, Takeru and Koyama, Masanori},
  booktitle={Proceedings of the 25th ACM SIGKDD international conference on knowledge discovery \& data mining},
  pages={2623--2631},
  year={2019}
}

@article{zaharia2018accelerating,
  title={Accelerating the machine learning lifecycle with MLflow.},
  author={Zaharia, Matei and Chen, Andrew and Davidson, Aaron and Ghodsi, Ali and Hong, Sue Ann and Konwinski, Andy and Murching, Siddharth and Nykodym, Tomas and Ogilvie, Paul and Parkhe, Mani and others},
  journal={IEEE Data Eng. Bull.},
  volume={41},
  number={4},
  pages={39--45},
  year={2018}
}

@book{lambert2018student,
  title={A student's guide to Bayesian statistics},
  author={Lambert, Ben},
  year={2018},
  publisher={SAGE Publications Ltd}
}

@book{bishop2023deep,
  title={Deep learning: Foundations and concepts},
  author={Bishop, Christopher M and Bishop, Hugh},
  year={2023},
  publisher={Springer Nature}
}

@article{rinaldi2024parameter,
  title={Parameter estimation from quantum-jump data using neural networks},
  author={Rinaldi, Enrico and Lastre, Manuel Gonz{\'a}lez and Herreros, Sergio Garc{\'\i}a and Ahmed, Shahnawaz and Khanahmadi, Maryam and Nori, Franco and Munoz, Carlos S{\'a}nchez},
  journal={Quantum Science and Technology},
  volume={9},
  number={3},
  pages={035018},
  year={2024},
  publisher={IOP Publishing}
}

@book{hastie2009elements,
  title={The elements of statistical learning: data mining, inference, and prediction},
  author={Hastie, Trevor and Tibshirani, Robert and Friedman, Jerome H and Friedman, Jerome H},
  volume={2},
  year={2009},
  publisher={Springer}
}

@article{ovadia2019can,
  title={Can you trust your model's uncertainty? evaluating predictive uncertainty under dataset shift},
  author={Ovadia, Yaniv and Fertig, Emily and Ren, Jie and Nado, Zachary and Sculley, David and Nowozin, Sebastian and Dillon, Joshua and Lakshminarayanan, Balaji and Snoek, Jasper},
  journal={Advances in neural information processing systems},
  volume={32},
  year={2019}
}

@article{lakshminarayanan2017simple,
  title={Simple and scalable predictive uncertainty estimation using deep ensembles},
  author={Lakshminarayanan, Balaji and Pritzel, Alexander and Blundell, Charles},
  journal={Advances in neural information processing systems},
  volume={30},
  year={2017}
}

@article{johansson2012qutip,
  title={QuTiP: An open-source Python framework for the dynamics of open quantum systems},
  author={Johansson, J Robert and Nation, Paul D and Nori, Franco},
  journal={Computer physics communications},
  volume={183},
  number={8},
  pages={1760--1772},
  year={2012},
  publisher={Elsevier}
}

@article{kiilerich2014estimation,
  title={Estimation of atomic interaction parameters by photon counting},
  author={Kiilerich, Alexander Holm and M{\o}lmer, Klaus},
  journal={Physical Review A},
  volume={89},
  number={5},
  pages={052110},
  year={2014},
  publisher={APS}
}

@book{murphy2023probabilistic,
  title={Probabilistic machine learning: Advanced topics},
  author={Murphy, Kevin P},
  year={2023},
  publisher={MIT press}
}

@article{zaheer2017deep,
  title={Deep sets},
  author={Zaheer, Manzil and Kottur, Satwik and Ravanbakhsh, Siamak and Poczos, Barnabas and Salakhutdinov, Russ R and Smola, Alexander J},
  journal={Advances in neural information processing systems},
  volume={30},
  year={2017}
}

@article{gammelmark2013bayesian,
  title={Bayesian parameter inference from continuously monitored quantum systems},
  author={Gammelmark, S{\o}ren and M{\o}lmer, Klaus},
  journal={Physical Review A—Atomic, Molecular, and Optical Physics},
  volume={87},
  number={3},
  pages={032115},
  year={2013},
  publisher={APS}
}

@article{nolan2021machine,
  title={A machine learning approach to Bayesian parameter estimation},
  author={Nolan, Samuel and Smerzi, Augusto and Pezz{\`e}, Luca},
  journal={npj Quantum Information},
  volume={7},
  number={1},
  pages={169},
  year={2021},
  publisher={Nature Publishing Group UK London}
}

@article{genois2021quantum,
  title={Quantum-tailored machine-learning characterization of a superconducting qubit},
  author={Genois, {\'E}lie and Gross, Jonathan A and Di Paolo, Agustin and Stevenson, Noah J and Koolstra, Gerwin and Hashim, Akel and Siddiqi, Irfan and Blais, Alexandre},
  journal={PRX Quantum},
  volume={2},
  number={4},
  pages={040355},
  year={2021},
  publisher={APS}
}

@book{quinonero2022dataset,
  title={Dataset shift in machine learning},
  author={Qui{\~n}onero-Candela, Joaquin and Sugiyama, Masashi and Schwaighofer, Anton and Lawrence, Neil D},
  year={2022},
  publisher={Mit Press}
}

@article{kiilerich2016bayesian,
  title={Bayesian parameter estimation by continuous homodyne detection},
  author={Kiilerich, Alexander Holm and M{\o}lmer, Klaus},
  journal={Physical Review A},
  volume={94},
  number={3},
  pages={032103},
  year={2016},
  publisher={APS}
}

@book{gelman1995bayesian,
  title={Bayesian data analysis},
  author={Gelman, Andrew and Carlin, John B and Stern, Hal S and Rubin, Donald B},
  year={1995},
  publisher={Chapman and Hall/CRC}
}

@article{cybenko1989approximation,
  title={Approximation by superpositions of a sigmoidal function},
  author={Cybenko, George},
  journal={Mathematics of control, signals and systems},
  volume={2},
  number={4},
  pages={303--314},
  year={1989},
  publisher={Springer}
}

@inproceedings{narang2017mixed,
  title={Mixed precision training},
  author={Narang, Sharan and Diamos, Gregory and Elsen, Erich and Micikevicius, Paulius and Alben, Jonah and Garcia, David and Ginsburg, Boris and Houston, Michael and Kuchaiev, Oleksii and Venkatesh, Ganesh and others},
  booktitle={Int. Conf. on Learning Representation},
  year={2017}
}

@article{paszke2019pytorch,
  title={Pytorch: An imperative style, high-performance deep learning library},
  author={Paszke, Adam and Gross, Sam and Massa, Francisco and Lerer, Adam and Bradbury, James and Chanan, Gregory and Killeen, Trevor and Lin, Zeming and Gimelshein, Natalia and Antiga, Luca and others},
  journal={Advances in neural information processing systems},
  volume={32},
  year={2019}
}

@article{lopez2022loss,
  title={Loss of antibunching},
  author={L{\'o}pez Carre{\~n}o, Juan Camilo and Zubizarreta Casalengua, Eduardo and Silva, Blanca and del Valle, Elena and Laussy, Fabrice P},
  journal={Physical Review A},
  volume={105},
  number={2},
  pages={023724},
  year={2022},
  publisher={APS}
}

@article{hoffman2014no,
  title={The No-U-Turn sampler: adaptively setting path lengths in Hamiltonian Monte Carlo.},
  author={Hoffman, Matthew D and Gelman, Andrew and others},
  journal={J. Mach. Learn. Res.},
  volume={15},
  number={1},
  pages={1593--1623},
  year={2014}
}

@article{feroz2008multimodal,
  title={Multimodal nested sampling: an efficient and robust alternative to Markov Chain Monte Carlo methods for astronomical data analyses},
  author={Feroz, Farhan and Hobson, Mike P},
  journal={Monthly Notices of the Royal Astronomical Society},
  volume={384},
  number={2},
  pages={449--463},
  year={2008},
  publisher={Blackwell Publishing Ltd Oxford, UK}
}

@article{buchner2023nested,
  title={Nested sampling methods},
  author={Buchner, Johannes},
  journal={Statistic Surveys},
  volume={17},
  pages={169--215},
  year={2023},
  publisher={The American Statistical Association, the Bernoulli Society, the Institute~…}
}

@book{wiseman2009quantum,
  title={Quantum measurement and control},
  author={Wiseman, Howard M and Milburn, Gerard J},
  year={2009},
  publisher={Cambridge university press}
}

@article{gawlikowski2023survey,
  title={A survey of uncertainty in deep neural networks},
  author={Gawlikowski, Jakob and Tassi, Cedrique Rovile Njieutcheu and Ali, Mohsin and Lee, Jongseok and Humt, Matthias and Feng, Jianxiang and Kruspe, Anna and Triebel, Rudolph and Jung, Peter and Roscher, Ribana and others},
  journal={Artificial Intelligence Review},
  volume={56},
  number={Suppl 1},
  pages={1513--1589},
  year={2023},
  publisher={Springer}
}

@article{aasi2013enhanced,
  title={Enhanced sensitivity of the LIGO gravitational wave detector by using squeezed states of light},
  author={Aasi, Junaid and Abadie, Joan and Abbott, BP and Abbott, Richard and Abbott, TD and Abernathy, MR and Adams, Carl and Adams, Thomas and Addesso, Paolo and Adhikari, RX and others},
  journal={Nature Photonics},
  volume={7},
  number={8},
  pages={613--619},
  year={2013},
  publisher={Nature Publishing Group UK London}
}

@article{backes2021quantum,
  title={A quantum enhanced search for dark matter axions},
  author={Backes, Kelly M and Palken, Daniel A and Kenany, S Al and Brubaker, Benjamin M and Cahn, SB and Droster, A and Hilton, Gene C and Ghosh, Sumita and Jackson, H and Lamoreaux, Steve K and others},
  journal={Nature},
  volume={590},
  number={7845},
  pages={238--242},
  year={2021},
  publisher={Nature Publishing Group UK London}
}

@book{kok2010introduction,
  title={Introduction to optical quantum information processing},
  author={Kok, Pieter and Lovett, Brendon W},
  year={2010},
  publisher={Cambridge university press}
}

@article{tse2019quantum,
  title={Quantum-enhanced advanced LIGO detectors in the era of gravitational-wave astronomy},
  author={Tse, Maggie and Yu, Haocun and Kijbunchoo, Nutsinee and Fernandez-Galiana, A and Dupej, P and Barsotti, L and Blair, CD and Brown, DD and Dwyer, SE ea and Effler, A and others},
  journal={Physical Review Letters},
  volume={123},
  number={23},
  pages={231107},
  year={2019},
  publisher={APS}
}

@article{cimini2023deep,
  title={Deep reinforcement learning for quantum multiparameter estimation},
  author={Cimini, Valeria and Valeri, Mauro and Polino, Emanuele and Piacentini, Simone and Ceccarelli, Francesco and Corrielli, Giacomo and Spagnolo, Nicol{\`o} and Osellame, Roberto and Sciarrino, Fabio},
  journal={Advanced Photonics},
  volume={5},
  number={1},
  pages={016005--016005},
  year={2023},
  publisher={Society of Photo-Optical Instrumentation Engineers}
}

@article{clark2025efficient,
  title={Efficient inference of quantum system parameters by approximate Bayesian computation},
  author={Clark, Lewis A and Ko{\l}ody{\'n}ski, Jan},
  journal={Physical Review Applied},
  volume={23},
  number={4},
  pages={044040},
  year={2025},
  publisher={APS}
}

@book{james2023statistical,
  title={An introduction to statistical learning: With applications in Python},
  author={James, Gareth and Witten, Daniela and Hastie, Trevor and Tibshirani, Robert and Taylor, Jonathan},
  booktitle={An introduction to statistical learning: With applications in Python},
  year={2023},
  publisher={Springer}
}

@article{khanahmadi2021time,
  title={Time-dependent atomic magnetometry with a recurrent neural network},
  author={Khanahmadi, Maryam and M{\o}lmer, Klaus},
  journal={Physical Review A},
  volume={103},
  number={3},
  pages={032406},
  year={2021},
  publisher={APS}
}

@article{greplova2017quantum,
  title={Quantum parameter estimation with a neural network},
  author={Greplova, Eliska and Andersen, Christian Kraglund and M{\o}lmer, Klaus},
  journal={arXiv preprint arXiv:1711.05238},
  year={2017}
}

@book{sisson2018handbook,
  title={Handbook of approximate Bayesian computation},
  author={Sisson, Scott A and Fan, Yanan and Beaumont, Mark},
  year={2018},
  publisher={CRC press}
}

@data{4AC59W_2025,
author = {Clark, Lewis A. and Kolodynski, Jan},
publisher = {Dane Badawcze UW},
title = "{Efficient inference of quantum system parameters by approximate Bayesian computation - ABC libraries}",
year = {2025},
version = {V2},
doi = {10.58132/4AC59W},
url = {https://doi.org/10.58132/4AC59W}
}

@dataset{sanchez_munoz_2023_8305509,
  author       = {Sánchez Muñoz, Carlos and
                  Rinaldi, Enrico},
  title        = {Dataset: Parameter estimation by learning quantum
                   correlations in continuous photon-counting data
                   using neural networks
                  },
  month        = oct,
  year         = 2023,
  publisher    = {Zenodo},
  version      = {0.0.0},
  doi          = {10.5281/zenodo.8305509},
  url          = {https://doi.org/10.5281/zenodo.8305509},
}

@book{murphy2022probabilistic,
  title={Probabilistic machine learning: an introduction},
  author={Murphy, Kevin P},
  year={2022},
  publisher={MIT press}
}

@article{yu2020quantum,
  title={Quantum correlations between light and the kilogram-mass mirrors of LIGO},
  author={Yu, Haocun and McCuller, L and Tse, M and Kijbunchoo, N and Barsotti, L and Mavalvala, N},
  journal={Nature},
  volume={583},
  number={7814},
  pages={43--47},
  year={2020},
  publisher={Nature Publishing Group UK London}
}

@article{brady2022entangled,
  title={Entangled sensor-networks for dark-matter searches},
  author={Brady, Anthony J and Gao, Christina and Harnik, Roni and Liu, Zhen and Zhang, Zheshen and Zhuang, Quntao},
  journal={PRX Quantum},
  volume={3},
  number={3},
  pages={030333},
  year={2022},
  publisher={APS}
}

@article{shi2023ultimate,
  title={Ultimate precision limit of noise sensing and dark matter search},
  author={Shi, Haowei and Zhuang, Quntao},
  journal={npj Quantum Information},
  volume={9},
  number={1},
  pages={27},
  year={2023},
  publisher={Nature Publishing Group UK London}
}

@book{devore2021modern,
  title={Modern mathematical statistics with applications},
  author={Devore, Jay L and Berk, Kenneth N and Carlton, Matthew A},
  year={2021},
  publisher={Springer Nature}
}

@dataset{anteneh_2025_17014659,
  author       = {Anteneh, Amanuel},
  title        = {Dataset: Parameter estimation with uncertainty
                   quantification from continuous measurement data
                   using neural network ensembles
                  },
  month        = sep,
  year         = 2025,
  publisher    = {Zenodo},
  doi          = {10.5281/zenodo.17014659},
  url          = {https://doi.org/10.5281/zenodo.17014659},
}

\end{document}